\documentclass[apj, twocolumn]{aastex631}
\usepackage{graphicx}
\usepackage{bm}
\usepackage{mathtools}
\usepackage{hyperref}
\usepackage{amsmath}
\usepackage{mathrsfs}
\usepackage{color}
\usepackage{lipsum}
\usepackage{soul}
\usepackage{comment}

\usepackage{hyperref}
\hypersetup{
 pdftitle={Outflow Properties Derived from a Broad Range of GRvRMHD Stellar Black Hole Accretion Disks}, 
 pdfauthor={Risti\'c}, 
 colorlinks=true,
 citecolor={blue},
}

\def\XCP{Computational Physics Division, Los Alamos National Laboratory, Los Alamos, NM 87545, USA}
\def\CTA{Center for Theoretical Astrophysics, Los Alamos National Laboratory, Los Alamos, NM 87545, USA}
\def\CAI{Computing and Artificial Intelligence Division, Los Alamos National Laboratory, Los Alamos, NM
 87545, USA}
\def\TD{Theoretical Division, Los Alamos National Laboratory, Los Alamos, NM 87545, USA}
\def\RIT{Center for Computational Relativity and Gravitation, Rochester Institute of Technology, Rochester, NY
 14623, USA}
\def\UCSD{Department of Astronomy and Astrophysics, University of California San Diego, La Jolla, CA 92093, USA}
\def\OR{Obsidian Research, Fort Wayne, IN 46835, USA}
\def\ND{Department of Physics and Astronomy, University of Notre Dame, Notre Dame, IN 46556, USA}
\def\SPARC{Michigan SPARC, Los Alamos National Laboratory, Ann Arbor, MI 48105, USA}

\graphicspath{{figures/}}

\begin{document}

\title{Outflow Properties from a Broad Range of GR$\nu$RMHD Stellar-Mass Black Hole Accretion Disks}

\author[0000-0001-7042-4472]{Marko~Risti\'c}
\affiliation{\CTA}
\affiliation{\TD}
\affiliation{\UCSD}
\email{markoris@ucsd.edu}
\author[0000-0001-6432-7860]{Jonah~M.~Miller}
\affiliation{\CTA}
\affiliation{\CAI}
\affiliation{\SPARC}
\author[0000-0001-5832-8517]{Richard~O'Shaughnessy}
\affiliation{\RIT}
\author[0000-0002-9950-9688]{Matthew~R.~Mumpower}
\affiliation{\CTA}
\affiliation{\CAI}
\affiliation{\ND}
\affiliation{\OR}
\author[0000-0003-1087-2964]{Christopher~J.~Fontes}
\affiliation{\CTA}
\affiliation{\XCP}
\author[0000-0003-2624-0056]{Chris~L.~Fryer}
\affiliation{\CTA}
\affiliation{\TD}
\author[0000-0003-3265-4079]{Ryan~T.~Wollaeger}
\affiliation{\CTA}
\affiliation{\CAI}

\date{\today}

\begin{abstract}
The majority of the mass ejected by neutron star mergers comes from accretion disk outflows. Over timescales of seconds, the thermodynamic evolution of the disk and its resultant outflows set the stage for the ejecta properties which manifest in kilonova observations at later times (hours/days).  Kilonova observables depend sensitively on key properties of these outflows --  mass ejection, the ejecta velocity distribution, late-time ejecta morphology and nucleosynthetic yields -- which can only be modeled with a detailed treatment of microphysical interactions including neutrino and general relativistic effects. In this work, we present a library of 25 general-relativistic magnetohydrodynamic accretion disks with neutrino transport, to our knowledge the largest to date, spanning a broad range of black hole and accretion disk combinations. From fits to our simulations, we derive analytic estimates of the average ejecta velocity distribution and mass ejection both at simulation end and extrapolated out to the end of the disk's lifetime. We also report ejecta morphology at timescales relevant to kilonova modeling, present spatially-dependent nucleosynthetic yields, and discuss the latter's dependence on initial disk conditions. While our disk library primarily includes initial conditions expected after neutron star mergers, a subset of our disks are applicable to other kilonova-adjacent scenarios. 
\end{abstract}

\section{Introduction}
\label{sec:intro}

The accretion disks formed around stellar mass black holes following a neutron star merger \citep{2013MNRAS.435..502F, 2014MNRAS.441.3444M, 2015MNRAS.448..541J, 2019MNRAS.482.3373F, 2019MNRAS.490.4811C, 2019PhRvD.100b3008M, 2024ApJ...962...79S, 2024ApJ...964..111L, 2025ApJ...985L...9L} or a collapsing, rapidly spinning massive star \citep[collapsar;][]{2019Natur.569..241S, 2020ApJ...902...66M} are hot ($T > 10^{9}$ K), dense $(\rho > 10^{10}$ g cm$^{-3})$, and neutron-rich $(n_n > 10^{32} \text{ cm}^{-3})$. These disks have been shown to be suitable environments for rapid neutron-capture process ($r$-process) nucleosynthesis \citep{2008ApJ...679L.117S, 2013MNRAS.435..502F}. As the majority of nuclei synthesized through the $r$-process are unstable, their radioactive decay heats the material ejected from the accretion disks, resulting in an optical/infrared transient called a kilonova \citep{2010MNRAS.406.2650M, Roberts2011, Goriely2011, 2012ApJ...746...48M, Korobkin_2012, 2021RvMP...93a5002C}. 
The multi-messenger detection of neutron star merger GW170817 \citep{LIGO-GW170817-bns, LIGO-GW170817-mma} and its associated kilonova AT2017gfo \citep{2017PASA...34...69A,2017Natur.551...64A,2017Sci...358.1556C,2017ApJ...848L..17C,Diaz17,2017Sci...358.1570D,Evans_2017,2017SciBu..62.1433H, 2017Sci...358.1559K, 2017ApJ...850L...1L,2017Natur.551...67P,2017Natur.551...75S,2017ApJ...848L..27T, 2017Natur.551...71T,2017PASJ...69..101U,2017ApJ...848L..24V, 2018ApJ...852L..30P} have provided a rich astrophysical probe for $r$-process nucleosynthesis. Initial studies of AT2017gfo have identified that fits to the photometric emission using semi-analytic kilonova models require the flux contribution from two ejecta components with distinct properties: a tidally ejected, heavy $r$-process dynamical ejecta and a (typically) light $r$-process, post-merger ejecta primarily driven by outflows from the resultant accretion disk \citep{2017ApJ...848L..17C}. Subsequent inferences using two-component radiative transfer simulations have found the post-merger ejecta component to be more massive than the dynamical ejecta component \citep{2022PhRvR...4a3046R, 2024PhRvR...6c3078P}, motivating a deeper dive into the diversity of post-merger ejecta properties that can arise from varied accretion disk realizations.

The impact of disk initial conditions on the production (or lack thereof) of heavy $r$-process elements in the disk, coupled with the direct influence of the disk outflows on subsequent kilonova signatures, warrants investigation into the ejecta and nucleosynthesis properties associated with disk outflows. These topics have been widely discussed in the literature to date; here, we present a non-exhaustive overview of recent works. \cite{2013MNRAS.435..502F} initially presented axisymmetric (two-dimensional) time-dependent hydrodynamic simulations of accretion disks with a neutrino treatment following the methodology of \cite{1985ApJS...58..771B}. Generally, they find relatively robust outflow compositions (characterized by electron fraction $Y_e = n_p/(n_p+n_n) \sim 0.20$), insensitive to black hole and disk initial conditions. They report that their outflow masses are weakly sensitive to the initial torus radius and their outflow radial velocities are remarkably robust with characteristic values $v_r \sim 0.067c$. In their follow-up study \citep{2014MNRAS.441.3444M}, they examine how the continuous neutrino irradiation from a long-lived hypermassive neutron star remnant can drive ejecta $Y_{\rm{e}}$ sufficiently high to suppress lanthanide production in the polar regions. This lower opacity ejecta permits brighter and bluer optical emission to punch through the polar regions before the redder infrared emission, thus providing a probe into the merger remnant's lifetime prior to collapse through observations of kilonovae. 

\cite{2015MNRAS.448..541J} subsequently presented black-hole torus systems simulated with a Newtonian hydrodynamics code including an energy-dependent, two-moment closure scheme for electron neutrino transport.\footnote{This two-moment scheme is commonly called ``M1" neutrino transport, corresponding to the closure of the Eddington tensor in the first moment of the Boltzmann transport equation.} Among many interesting results, they observe an inverse relationship between the strength of the third abundance peak and black hole mass, and generally find that nucleosynthesis is more sensitive to black hole parameters than to the torus parameters, specifically torus mass. \cite{2018ApJ...860...64F} present a detailed study using axisymmetric general relativistic neutrino radiation-hydrodynamics (GR$\nu$RHD) simulations to explore a torus surrounding a massive neutron star remnant, corroborating the results from \cite{2014MNRAS.441.3444M} that the remnant's neutrino irradiation of the ejecta drives $Y_e \geq 0.25$ and suppresses lanthanide production, producing a rapidly evolving, bright, and blue transient, under the assumption of little polar dynamical ejecta. \cite{2018ApJ...858...52S} conversely present the first three-dimensional GR$\nu$RMHD post-merger disk simulations, showing that the disk maintains a low $Y_e \lesssim 0.1$ in the equatorial midplane, providing a neutron reservoir sufficiently rich to produce third-peak r-process elements in the disk outflow ejecta ($Y_e \sim 0.2$). Shortly thereafter, \cite{2019Natur.569..241S} presented three-dimensional general-relativistic magnetohydrodynamic (GRMHD) simulations using approximate neutrino transport via a leakage scheme, suggesting that collapsars may be the dominant sources of $r$-process in the universe, contributing more than $80\%$ of the total $r$-process material. In the same year, \cite{2019MNRAS.482.3373F} and \cite{2019MNRAS.490.4811C} performed long ($\sim$9 s and $\sim$4 s, respectively) GRMHD simulations with parametrized neutrino effects, showing large $\sim 40\%$ disk mass ejection and higher average velocities while matching the red kilonova emission from AT2017gfo but underpredicting blue emission. Motivated by the discovery of a black hole-neutron star merger with no associated electromagnetic counterpart, \cite{2020MNRAS.497.3221F} followed up with a study of the landscape of disk ejecta for such systems. They presented a large library of axisymmetric, long-term ($\mathcal{O}(10)$ s) hydrodynamic
simulations of the viscous evolution of disks surrounding stellar-mass black holes; their results indicate that more massive disks eject a smaller fraction of their mass such that, for $M_{\rm{ej}} \propto M_{\rm{disk}}^p$, $p < 1$. Finally, we highlight the works by \cite{2019PhRvD.100b3008M, 2020ApJ...902...66M, 2023ApJ...945L..13C, 2024ApJ...962...79S, 2024ApJ...964..111L, 2025ApJ...985L...9L} which use $\nu$bhlight, the code employed in this work, and discuss their contributions in Section~\ref{sec:results}. We refer the reader to the review by \cite{2019ARNPS..69...41S} and a handful of the many works published since \citep{2022PhRvD.106j3003F, 2022PhRvD.105h3024J, 2023PhRvL.131a1401K, 2025ApJ...984...77G, 2025ApJ...985L..26I, 2025PhRvD.111b3015K, 2025PhRvD.111h3042S}.

Although the aforementioned studies all examine stellar mass black hole accretion disk outflows, each employs a slightly different computational technique to tackle different physical phenomena. Our code, \texttt{$\nu$bhlight} \citep{nubhlight}, is capable of producing GRMHD simulations in three dimensions with Boltzmann neutrino transport, offering one of the most physically motivated representations of the microphysics in these disks (though at the cost of the detailed late-time mass ejection information afforded by axisymmetric simulations).
In this study, we complement these aforementioned works by presenting the largest simulation library to date which encompasses a broad range of GR$\nu$RMHD accretion disks surrounding stellar mass black holes. 
We use the uncertainty-optimized simulation placement scheme described in \cite{2022PhRvR...4a3046R} and \cite{2025arXiv250903003R} to generate a variety of scenarios representative of both neutron star merger and collapsar environments. In conjunction with simulations presented in previous studies, we compile a library of disk simulations and report on general trends such as mass ejection and ejecta radial velocity distributions. We present nucleosynthetic yields for a handful of special cases, selected based on ejecta velocity and viewing angle cuts, which are relevant to downstream kilonova simulations. While not all of the disks in our study are likely to form in an astrophysical scenario, understanding the thermodynamic evolution of a variety of disks is essential to building intuition about their behavior around both stellar and supermassive black holes. We publish our ejecta tracers on a publicly available Zenodo repository \citep{zenodo} in a condensed format due to data volume restrictions.

The paper is organized as follows. Section~\ref{sec:methods} describes our \texttt{$\nu$bhlight} GR$\nu$RMHD disk simulations, simulation placement strategy informed by Gaussian process variance, ejecta extrapolation techniques, and \texttt{PRISM} nuclear reaction network simulations. In Section~\ref{sec:results}, we show the most relevant results pertaining to our disk simulations, alongside fits to the radial velocity distributions and ejected mass. We also present nucleosynthesis yields and their associated mass histograms in $Y_{\rm{e}}, \theta$ space for each disk. In Section~\ref{sec:discussion}, we discuss the implications of our results and conclude in Section~\ref{sec:conclusion}. 

\section{Methods}
\label{sec:methods}

\subsection{Disk Simulations with \texorpdfstring{$\nu$}bbhlight}
\label{sec:disks}

A thorough description of the \texttt{$\nu$bhlight} simulation methodology, including detailed code verification, is presented in \cite{2019ApJS..241...30M}. For brevity, here we discuss only the portions most relevant to the details of our study. Throughout this work, we assume geometric units in which $G = c = 1$.

\texttt{$\nu$bhlight} solves the general-relativistic magnetohydrodynamic equations with neutrino transport in stationary spacetimes. Fluid integration is performed with a second-order shock-capturing scheme using the finite volume method implemented in the High-Accuracy Relativistic Magnetohydrodynamics (\texttt{HARM}) code \citep{2003ApJ...589..444G}.  Frequency-dependent radiation transport is performed with the second-order Monte Carlo scheme from \texttt{grmonty} \citep{2009ApJS..184..387D}. Fluid and radiation exchange four-momentum in an explicit first-order operator-split fashion, coupled together using \texttt{bhlight} \citep{2015ApJ...807...31R}. \texttt{$\nu$bhlight} tracks the evolution of relativistic neutrinos while conserving lepton number, energy and momentum. Thermodynamic quantities are additionally tracked using ${\sim} 10^6$ tracer particles which are passively advected with the fluid, of which $\mathcal{O}(10^5)$ become unbound; tracers are considered unbound once they cross the $r=250M_{\rm{bh}}$ surface. These tracers are sampled roughly uniformly across grid cells which contain physical mass (i.e., not the artificial atmosphere), with the total mass in a finite-volume cell being distributed among the N uniformly sampled tracer particles in that cell. We use a radially logarithmic, quasi-spherical grid in $(r, \theta, \phi)$ with $N_r \times N_\theta \times N_\phi$ = $192 \times 168 \times 66$ \citep{2003ApJ...589..444G}. Each simulation runs for $t = 10^4 M_{\rm{bh}}$, corresponding to ${\sim}5 \times 10^3$ light-crossing times across the black hole horizon.

Each simulation assumes a disk surrounding a black hole in a stationary Kerr spacetime \citep{1963PhRvL..11..237K}. The disk starts as a Fishbone-Moncrief torus in hydrostatic equilibrium \citep{FM76} assuming a constant specific angular momentum, entropy, electron fraction, and total mass. Each torus is initialized with a single poloidal magnetic field loop with a minimum ratio of gas to magnetic pressure of 100. As the disk evolves, magnetorotational instability \citep{1991ApJ...376..214B} develops and shifts the disk to a turbulent state, providing the necessary turbulent viscosity for accretion onto the black hole \citep{1973A&A....24..337S}. We assume the SFHo equation of state \citep{SteinerSFHo} as provided online in tabular form\footnote{https://stellarcollapse.org/equationofstate.html}.
The \texttt{torus\_id.py} script used to generate the tori can be found as part of the \texttt{$\nu$bhlight} GitHub repository \citep{nubhlight}.

Each \texttt{$\nu$bhlight} simulation is characterized by black hole and accretion disk parameters $\vec{x} = [M_{\rm{bh}}, a_{\rm{bh}}, M_{\rm{disk}}, Y_{\rm{e}}, s]$ and the associated ejecta outflow parameters $\vec{y}(\theta) = [M_{\rm{ej}}, M_{\rm{ej}}/M_{\rm{acc}}, \langle Y_{\rm{e}} \rangle, M_{\rm{ej, ext}}]$,  
where the elements of $\vec{x}$ correspond to black hole mass, black hole spin, initial disk mass, initial disk $Y_{\rm{e}}$, and initial entropy, and the outputs $\vec{y}$ to the ejected mass, the ratio of ejected to accreted mass, the mass-averaged $Y_{\rm{e}}$, and ejected mass extrapolated to late times as a function of angle. For the full training data, we consider the total, equatorial $(|\theta| \geq 75^\circ)$, and polar $(0^\circ \leq |\theta| \leq 40^\circ)$ outflows, with $\theta$ defined as the polar angle ($\theta = 0$ along the spin axis), for a total of 12 training outputs (4 each for total, equatorial, and polar ejecta cases). Masses are in units of solar masses $M_\odot$ and entropies in units of $k_{\rm{B}}/\rm{baryon}$, while all remaining quantities are dimensionless. The extrapolated ejected mass $M_{\rm{ej, ext}}$ assumes mass ejection continues as the ratio $M_{\rm{ej}} / M_{\rm{acc}}$ until the accretion disk mass is exhausted (though we do refer the reader to Figure~2 of \cite{2024ApJ...962...79S} which clearly shows that this is not necessarily a good assumption at times $\gtrsim1$ s). We note that the $\theta$ cuts for the equatorial and polar regions \emph{are only made for the purposes of constructing our training dataset}. For the training data only, they do not necessarily encompass all simulation tracers as the angles $40^\circ < |\theta| < 75^\circ$ are omitted for these specific cuts such that $M_{\rm{ej, eq}} + M_{\rm{ej, pol}} \not\equiv M_{\rm{ej}}$. This choice is made to preferentially guide simulation placement toward regions of interest for kilonova modeling, which are typically modeled as an equatorial dynamical ejecta and a polar secular ejecta. However, we emphasize that in every \texttt{$\nu$bhlight} simulation, the tracers equally sample all angles $\theta$, and the outflow results and analysis presented later in this work encompass all $\theta$ unless specified otherwise.

\subsection{Simulation Placement}
\label{sec:placement}

\begin{figure}[t]
\includegraphics[width=\linewidth]{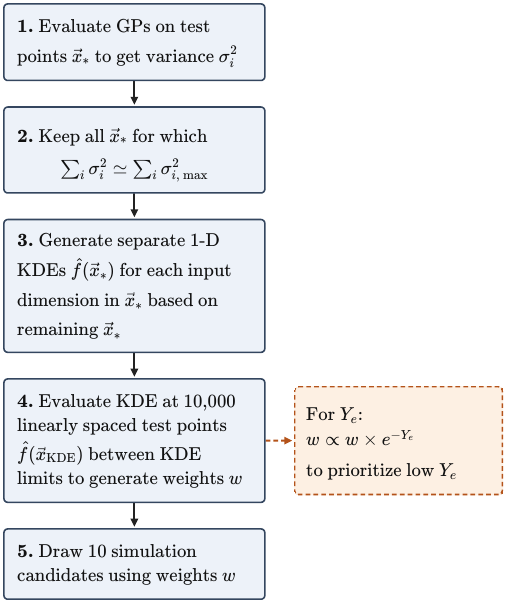}
\caption{Algorithm design flowchart describing Gaussian process (GP), Gaussian kernel density estimate (KDE), and simulation candidate sampling during the simulation placement process. The weights $w$ for $Y_{\rm{e}}$ are additionally modified to prioritize lower $Y_{\rm{e}}$ values, as might be expected in a newly formed accretion disk.}
\label{fig:flowchart}
\end{figure}

Our optimized simulation placement is guided by Gaussian process (GP) \citep{RW_GP} interpolation uncertainty given the training inputs $\vec{x}$ and outputs $\vec{y}$ described in Section~\ref{sec:disks}. We construct each GP in \texttt{scikit-learn} \citep{scikit-learn} using a white noise kernel added to the product of a constant and radial basis function kernel with all kernel hyperparameters initialized using the \texttt{scikit-learn} default values. We train an individual GP on each of the outflow parameters in $\vec{y}$, resulting in 12 separate GPs, each evaluating a scalar prediction of the relevant outflow quantity as a function of the 5-dimensional input $\vec{x}$. From this point, we follow the algorithm outlined in Figure~\ref{fig:flowchart} to generate simulation placement candidates based on GP variance maximization.

To understand the variance landscape for each outflow quantity, we evaluate each GP on 1,000 test points $\vec{x}_*$. These test points are drawn uniformly between the minima and maxima of each parameter from the initial grid (i.e., the bolded simulations in Tables~\ref{tbl:disks} and \ref{tbl:ejecta}). The sole exception is black hole spin, for which we impose a lower bound of $0.6 < a_{\rm{bh}}$ to prioritize high-spin black holes which provide more realistic initial conditions following a merger or collapsar event, e.g., the near-universal post-merger black hole spin around $a \sim 0.7$ \citep{2008ApJ...684..822B}.
For each test point $\vec{x}_{*, k}$, we are only interested in its associated variance $\sigma_i^2$ from the $i$-th GP.
We leverage a behavior exhibited by unconverged GPs in the small training dataset limit where test inputs $\vec{x}_*$ far from those in the training data $\vec{x}$ return a consistent maximum value for the variance $\sigma^2_{i,\rm{max}}$. The exact value of $\sigma_{\rm{i,max}}$ is dependent on specific choices of the kernel and its parameters and is different for each GP used here. It follows, then, that the test points in the most unconstrained regions of parameter space will reach this maximum uncertainty across all GPs such that

\begin{equation}
\sum_i \sigma_i^2 \simeq \sum_i \sigma_{i, {\rm{max}}}^2.
\end{equation}

For this application, we strictly focus on the $\sigma_{\rm{max}}^2$ behavior and not the physical interpretation of the GP variance given that some GP variances carry units while others are dimensionless. Using this property, we prioritize unexplored regions of parameter space with the following criterion:

\begin{equation}
\vec{x_{*,k}} = 
\begin{cases}
    \text{keep} & \text{if } \sum_i \sigma_i^2 \simeq \sum_i \sigma_{i, {\rm{max}}}^2 \\
    \text{discard}  & \text{if } \sum_i \sigma_i^2 < \sum_i \sigma_{i, {\rm{max}}}^2,
\end{cases}
\label{eq:cut}
\end{equation}

where we employ an approximate equality for the keep case to account for numerical rounding. We caution readers that, in this instance, we have not normalized the scales for individual $\sigma_{i, \rm{max}}^2$, and are therefore susceptible to an individual feature dominating simulation placement.

We deliberately only draw $N = 1,000$ test points $\vec{x}_*$ to circumvent the unfavorable GP variance prediction calculation which scales as $\mathcal{O}(N^2)$. Instead, using the $\vec{x}_*$ test points which pass the criterion in Equation~\ref{eq:cut}, we generate a one-dimensional Gaussian kernel density estimate (KDE) $\hat{f}_d(\vec{x_*})$ for each input dimension $d$ in $\vec{x}$ as a proxy for the probability density function of the maximum-variance test points. Each KDE is bounded between $(\min\vec{x}_{*, d}, \max \vec{x}_{*, d})$ for each dimension $d$ in the inputs. It warrants mentioning that assuming the black hole and accretion disk parameters are separable in this way is a strong assumption; however, later discussion in Section~\ref{sec:discussion} will show that the consequences of the simulation placement methodology may be subdominant to the stability of the Fishbone-Moncrief torus for a given set of initial conditions.

We sample 10,000 linearly spaced points $\vec{x}_{\rm{KDE}}$ between the limits of each KDE and evaluate them to generate weights $w = \hat{f}_d(x_{\rm{KDE}})$. For the disk mass, we adjust the KDE bounds to the remnant baryon mass prediction (based on black hole mass, spin and neutron star compactness) from \cite{Foucart18}, accounting for their baryonic mass normalization using Equation 36 from \cite{LP01}. Additionally, we scale the weights $w$ for the electron fraction proportional to $\exp{(-Y_e)}$ such that, for the initial disk $Y_{\rm{e}}$ dimension exclusively, the weights are $w \rightarrow w \times \exp{(-Y_e})$. This additional weight factor results in values around $Y_e \sim 0.10$ being about twice as likely to be sampled as $Y_e \sim 0.36$. 

Finally, we randomly draw 10 candidate simulation parameter combinations from the KDEs using the weights $w$. We caution that the choice of the 10 proposed simulation parameters is entirely guided by the algorithm described above and in Figure~\ref{fig:flowchart} rather than physically intuitive or realistic choices. As such, many of the Fishbone-Moncrief tori related to these potentially unrealistic or unstable disks were not realized; the final simulation grid suffers from survivorship bias related to successfully constructed and simulated Fishbone-Moncrief tori (as evident from, e.g., the lack of very low $Y_{\rm{e}}$ GP-placed disks despite the preferential low-$Y_{\rm{e}}$ weighting).

\subsection{Ejecta Extrapolation}
\label{sec:mej_ext}

With our simulations terminating at $t = 10^4 M_{\rm{bh}}$, we evolve the disk out to a maximum time of ${\sim} 360$ ms in the case of our heaviest black hole with $M=10 M_\odot$. As shown in previous work \citep[e.g.,][]{2019MNRAS.482.3373F}, these disks can continue evolving for many seconds. As such, our simulations are not representative of a complete disk evolution (where complete means the point at which the disk has fully exhausted its matter content, either through ejection or accretion). To understand the disk ejecta relevant to downstream kilonova modeling, we also calculate extrapolated ejecta and accretion values based on the disk masses and accretion rates at the end of each simulation. 

To account for the tracer travel time delay to our extraction surface $r = 250M_{\rm{bh}}$, we calculate the time shift $t_{\rm{shift}} = 250M/v_{\rm{max}}$. Here, $v_{\rm{max}}$ is in units of $c$ and is chosen as the $v_r$ distribution's $3\sigma$ upper bound, i.e., the 99.73 percentile for the average skew-normal distribution discussed in Section~\ref{sec:vr}. The choice of $v_{\rm{max}}$ as the $3\sigma$ upper bound rather than the distribution maximum omits fast tracer outliers that would disproportionately underestimate $t_{\rm{shift}}$. This approach also assumes purely radial motion from the origin, which black hole spin alone rules out. We emphasize that by applying $t_{\rm{shift}}$ to binned tracers rather than each tracer individually and by using only the radial tracer velocity at $r = 250M_{\rm{bh}}$ instead of each tracer's velocity history along its trajectory, we are prone to unphysical time shifts in some cases (see Section~\ref{sec:discussion}). 

We calculate our extrapolated mass accretion $\dot{M}_{\rm{bh, ext}}$ and ejection $\dot{M}_{\rm{ext}}$ rates by linearly fitting the black hole accretion $\dot{M}_{\rm{bh}}$ and disk ejection $\dot{M}$ rates for $t \geq 5000M_{\rm{bh}}$. At these simulation times, both accretion and ejection have reached mostly stable rates, allowing for reasonable power-law extrapolation. We assume that $\dot{M}$ at simulation end (the starting point of $\dot{M}_{\rm{ext}}$) corresponds to $t = 10^4M - t_{\rm{shift}}$. To calculate the disk lifetime $\tau_{\rm{ext}}$, we iteratively subtract mass lost to accretion and ejection from the remaining disk mass $M_{\rm{disk}}$ in time steps of $\Delta t = 1 \text{ ms} \sim 70M_{\rm{bh}}$ until $M_{\rm{disk}} = 0$. For cases where both $\dot{M}_{\rm{bh, ext}}$ and $\dot{M}_{\rm{ext}}$ are negatively sloped (i.e., do not lead to the convergence of $M_{\rm{disk}}$ toward 0 as $t$ increases), we assume their values are constant and match those at simulation end for the entire extrapolation process. 

\subsection{Nucleosynthesis Simulations with PRISM}
\label{sec:prism}

We perform nucleosynthesis calculations with version 1.6.0 of the Portable Routines for Integrated nucleoSynthesis Modeling (PRISM) nuclear reaction network \citep{Sprouse2021}.
As each disk produces $\mathcal{O}(10^5)$ unbound tracers, we downselect the nucleosynthesis tracers using a similar scheme to the one presented in \cite{2025ApJ...985L...9L}.
First, we ensure that all nucleosynthesis candidate tracers satisfy the Bernoulli criterion. This criterion suggests, though does not guarantee, that the tracers are on a permanently unbound trajectory; by selecting tracers which must pass an extraction surface, we also miss the population of tracers which are not yet unbound, but would become so at later simulation times (see, e.g., \cite{2021PhRvD.104l3010F} for a more detailed discussion about ejecta criteria and possible effects on nucleosynthesis).
We separate the tracers into 15 equally spaced bins across $0 \leq Y_e \leq 0.5$, as in \cite{2025ApJ...985L...9L}.
Unlike their approach, we cannot necessarily assume that each disk will have a low- and a high-velocity tracer in each $Y_{\rm{e}}$ bin; instead, we perform one additional cut at a time for each batch of nucleosynthesis runs. Specifically, we examine the separate cases where $v_r < 0.08c$, $v_r \geq 0.08c$, $\theta < 30^\circ$, and $\theta \geq 30^\circ$.
Each of these cases assumes 15 $Y_{\rm{e}}$ bins, though tracers are not guaranteed to exist in each bin; in fact, most disks only have tracers in a few bins, with some extreme cases having tracers only in the highest $Y_{\rm{e}}$ bin (Disk 24).
Each of the selected tracers assumes the mass of all tracers in the given $Y_{\rm{e}}$ bin such that the final nucleosynthetic yields are mass-weighted according to each $Y_{\rm{e}}$ bin's relative contribution to the total mass.
Once the nucleosynthesis tracers are selected, their trajectories are filtered to include only information past the last time at which the temperature crosses $10 \pm 0.1$ gigakelvin (GK). 
This filtering ensures that the tracer does not return to the disk and re-experience nuclear statistical equilibrium (NSE) past this point, but rather smoothly continues on a cooling trajectory, ensuring continuous nucleosynthesis without complex behavior brought on by reheating into NSE.
The filtered trajectories are provided to the PRISM nuclear reaction network, producing nucleosynthesis yields for each $Y_{\rm{e}}$ bin. 
The final nucleosynthesis yields for each disk are mass-weighted averages using the mass of each $Y_{\rm{e}}$ bin as a weight, as described above.

\section{Results}
\label{sec:results}

\begin{deluxetable*}{cccccc|ccccccc}
\centering
\tablewidth{0pt}
\tablecolumns{13}
\tablehead{
\colhead{Disk \#} & \colhead{$M_{\rm{bh}}$} & \colhead{$a_{\rm{bh}}$} & \colhead{$M_{\rm{disk}}$} & \colhead{$Y_{\rm{e}}$} & \colhead{$s$} \vrule & \colhead{$\dot{M}_{\rm{peak}}$} & \colhead{$\langle T \rangle$} & \colhead{$\langle \rho \rangle$} & \colhead{$P_{\rm{jet, max}}$} & \colhead{$L_{\nu, \rm{peak}}$} & \colhead{$\alpha$} & \colhead{$\tau$}\\
\colhead{--} & \colhead{[$M_\odot$]} & \colhead{--} & \colhead{[$M_\odot$]} & \colhead{--} & \colhead{$ \left[ \frac{k_{\rm{b}}}{\rm{baryon}} \right] $} \vrule & \colhead{[$M_\odot$ s$^{-1}$]} & \colhead{[GK]} & \colhead{[g cm$^{-3}$]} & \colhead{[erg s$^{-1}$]} & \colhead{[erg s$^{-1}$]} & \colhead{--} & \colhead{[sec]}}
\vspace{5mm}
\startdata
\textbf{1} & \textbf{2.58} & \textbf{0.00} & \textbf{0.12} & \textbf{0.10} & \textbf{4.00} & \textbf{1.506} & \textbf{15.11} & \textbf{8.8e+10} & \textbf{0 ($a_{\rm{bh}} = 0$)} & \textbf{1.4e+52} & \textbf{0.384} & \textbf{1.66} \\
\textbf{\textit{2}} & \textbf{\textit{2.58}} & \textbf{\textit{0.69}} & \textbf{\textit{0.12}} & \textbf{\textit{0.10}} & \textbf{\textit{4.00}} & \textbf{\textit{1.831}} & \textbf{\textit{21.10}} & \textbf{\textit{5.3e+10}} & \textbf{\textit{4.7e+51}} & \textbf{\textit{3.9e+52}} & \textbf{\textit{0.103}} & \textbf{\textit{0.58}} \\
\textbf{3} & \textbf{3.00} & \textbf{0.80} & \textbf{0.0002} & \textbf{0.50} & \textbf{8.00} & \textbf{0.002} & \textbf{1.84} & \textbf{4.9e+06} & \textbf{8.4e+48} & \textbf{3.3e+48} & \textbf{0.076} & \textbf{0.16} \\
\textbf{4} & \textbf{3.00} & \textbf{0.80} & \textbf{0.003} & \textbf{0.10} & \textbf{8.00} & \textbf{0.033} & \textbf{12.94} & \textbf{1.2e+09} & \textbf{1.5e+50} & \textbf{7.8e+50} & \textbf{0.102} & \textbf{0.37} \\
\textbf{5} & \textbf{3.00} & \textbf{0.80} & \textbf{0.0016} & \textbf{0.50} & \textbf{8.00} & \textbf{0.014} & \textbf{5.75} & \textbf{1.9e+08} & \textbf{8.6e+49} & \textbf{3.7e+50} & \textbf{0.136} & \textbf{0.38} \\
\textbf{6} & \textbf{3.00} & \textbf{0.80} & \textbf{0.02} & \textbf{0.10} & \textbf{8.00} & \textbf{0.135} & \textbf{16.82} & \textbf{4.1e+09} & \textbf{4.6e+50} & \textbf{4.3e+51} & \textbf{0.126} & \textbf{0.29} \\
\textbf{\textit{7}} & \textbf{\textit{3.00}} & \textbf{\textit{0.80}} & \textbf{\textit{0.02}} & \textbf{\textit{0.50}} & \textbf{\textit{8.00}} & \textbf{\textit{0.134}} & \textbf{\textit{14.99}} & \textbf{\textit{3.8e+09}} & \textbf{\textit{9.1e+50}} & \textbf{\textit{3.2e+52}} & \textbf{\textit{0.127}} & \textbf{\textit{0.35}} \\
8  & 3.90 & 0.64 & 0.289 & 0.34 & 6.56 & 0.707 & 5.82 & 6.5e+08 & 3.2e+51 & 9.4e+53 & 0.140 & \hspace{4mm} 0.40 (*) \\
\textbf{\textit{9}} & \textbf{\textit{5.00}} & \textbf{\textit{0.90}} & \textbf{\textit{0.42}} & \textbf{\textit{0.10}} & \textbf{\textit{4.00}} & \textbf{\textit{3.088}} & \textbf{\textit{22.56}} & \textbf{\textit{5.0e+10}} & \textbf{\textit{2.2e+52}} & \textbf{\textit{1.3e+53}} & \textbf{\textit{0.185}} & \textbf{\textit{1.19}} \\
10  & 5.25 & 0.61 & 0.183 & 0.39 & 7.57 & 0.499 & 12.15 & 3.4e+09 & 1.3e+51 & 9.4e+52 & 0.011 & \hspace{4mm} 2.18 (\textdagger) \\
11  & 5.59 & 0.61 & 0.231 & 0.37 & 6.96 & 0.594 & 6.92 & 3.9e+08 & 1.8e+51 & 1.3e+53 & 0.062 & 3.19 \\
12  & 5.68 & 0.84 & 0.196 & 0.44 & 4.65 & 0.706 & 7.61 & 1.6e+09 & 3.1e+51 & 3.0e+53 & 0.233 & \hspace{4mm} 2.45 (*) \\
13  & 5.96 & 0.79 & 0.131 & 0.27 & 6.73 & 0.689 & 14.51 & 7.8e+09 & 3.9e+51 & 1.5e+53 & 0.059 & 3.63 \\
\textbf{\textit{14}} & \textbf{\textit{6.00}} & \textbf{\textit{0.75}} & \textbf{\textit{0.143}} & \textbf{\textit{0.15}} & \textbf{\textit{8.00}} & \textbf{\textit{0.688}} & \textbf{\textit{17.76}} & \textbf{\textit{7.4e+09}} & \textbf{\textit{2.3e+51}} & \textbf{\textit{3.2e+52}} & \textbf{\textit{0.147}} & \textbf{\textit{5.23}} \\
15  & 6.25 & 0.82 & 0.199 & 0.31 & 4.29 & 1.102 & 12.83 & 1.2e+10 & 4.3e+51 & 7.9e+52 & 0.024 & \hspace{4mm} 1.29 (*\textdagger) \\
16  & 6.37 & 0.78 & 0.133 & 0.48 & 6.02 & 0.470 & 6.53 & 7.6e+08 & 1.8e+51 & 1.2e+53 & 0.023 & \hspace{4mm} 1.71 (\textdagger) \\
\textbf{\textit{17}} & \textbf{\textit{7.00}} & \textbf{\textit{0.90}} & \textbf{\textit{0.25}} & \textbf{\textit{0.10}} & \textbf{\textit{4.00}} & \textbf{\textit{2.376}} & \textbf{\textit{17.33}} & \textbf{\textit{1.7e+10}} & \textbf{\textit{7.3e+51}} & \textbf{\textit{8.6e+52}} & \textbf{\textit{0.071}} & \textbf{\textit{1.64}} \\
18  & 8.13 & 0.71 & 0.044 & 0.35 & 5.07 & 0.166 & 8.16 & 1.2e+09 & 3.2e+50 & 2.9e+51 & 0.085 & 4.07 \\
19  & 8.33 & 0.66 & 0.057 & 0.43 & 7.43 & 0.125 & 3.51 & 5.5e+07 & 3.0e+50 & 4.7e+51 & 0.073 & 1.09 \\
20  & 8.48 & 0.77 & 0.114 & 0.34 & 6.27 & 0.419 & 10.99 & 2.5e+09 & 1.6e+51 & 3.1e+52 & 0.172 & 1.31 \\
21  & 9.08 & 0.77 & 0.072 & 0.33 & 6.40 & 0.413 & 9.08 & 1.2e+09 & 9.6e+50 & 1.0e+52 & 0.167 & 1.94 \\
22  & 9.38 & 0.79 & 0.064 & 0.41 & 5.22 & 0.217 & 7.01 & 3.4e+08 & 4.9e+50 & 4.5e+51 & 0.113 & 2.19 \\
23  & 9.68 & 0.74 & 0.109 & 0.42 & 5.66 & 0.289 & 5.58 & 3.7e+08 & 6.6e+50 & 1.1e+52 & 0.130 & 2.53 \\
24  & 9.85 & 0.68 & 0.098 & 0.49 & 4.05 & 0.271 & 8.01 & 1.2e+09 & 3.8e+50 & 4.9e+51 & 0.118 & 6.49 \\
\textbf{\textit{25}} & \textbf{\textit{10.00}} & \textbf{\textit{0.80}} & \textbf{\textit{0.082}} & \textbf{\textit{0.20}} & \textbf{\textit{8.00}} & \textbf{\textit{0.393}} & \textbf{\textit{9.18}} & \textbf{\textit{5.7e+08}} & \textbf{\textit{1.3e+51}} & \textbf{\textit{1.6e+52}} & \textbf{\textit{0.093}} & \hspace{4mm} \textbf{\textit{0.87}} (*) \\
\hline
Mean & -- & -- & -- & -- & -- & 0.67 & 10.96 & 1.0e+10 & 2.5e+51 & 9.0e+52 & 0.118 & 1.89 \\
No outliers & -- & -- & -- & -- & -- & 0.63 & 11.56 & 1.0e+10 & 1.8e+51 & 5.7e+52 & 0.120 & 2.17 \\
Median & -- & -- & -- & -- & -- & 0.42 & 9.18 & 1.2e+09 & 1.3e+51 & 3.1e+52 & 0.113 & 1.64 \\
No outliers & -- & -- & -- & -- & -- & 0.44 & 11.57 & 2.0e+09 & 1.1e+51 & 3.2e+52 & 0.115 & 1.82 \\
\enddata
\label{tbl:disks}
\caption{Selected disk parameters of interest for the initial black hole and accretion disk parameters given on the left-hand side of the table. Simulation input parameters are black hole mass $M_{\rm{bh}}$, black hole spin $a_{\rm{bh}}$, initial disk mass $M_{\rm{disk}}$, initial disk $Y_{\rm {e}}$, and initial disk entropy $s$. Disk parameters include the peak accretion rate $\dot{M}_{\rm{peak}}$, mass-weighted temperature at peak accretion rate $\langle T\rangle$, mass-weighted density at peak accretion $\langle \rho \rangle$, the maximum jet power at the time of peak accretion $P_{\rm{jet, max}}$, maximum neutrino luminosity at peak accretion $L_{\nu, {\rm{max}}}$, the dimensionless Shakura--Sunyaev effective viscosity parameter $\alpha$ evaluated at the time of peak accretion and at the radius of peak pressure, and the extrapolated disk lifetime $\tau$. Simulations which terminate prior to their expected number of orbits have an asterisk $(*)$ appended to the end of their respective row, while those which have been reduced to axisymmetry (2D) are denoted with a dagger (\textdagger). Bolded entries are used as the initial training set for the Gaussian process placement scheme, while italicized entries have previously appeared in the literature. The bottom four rows represent the mean values across all disks for the selected quantities (first row), the mean values for disks which are not $v_r$ distribution outliers (second row), with the same structure for the median values (third and fourth rows).}
\end{deluxetable*}

\begin{deluxetable*}{cccccc|cccccccc}
\centering
\tablewidth{0pt}
\tablecolumns{14}
\tablehead{
\colhead{Disk \#} & \colhead{$M_{\rm{bh}}$} & \colhead{$a_{\rm{bh}}$} & \colhead{$M_{\rm{disk}}$} & \colhead{$Y_{\rm{e}}$} & \colhead{$s$} \vrule & \colhead{$\langle v_r \rangle$} & \colhead{$\langle s \rangle$} & \colhead{$\langle Y_{\rm{e}} \rangle$} & \colhead{$X_{\rm{La}}$} & \colhead{$X_{\rm{Ac}}$} & \colhead{$M_{\rm{ej}}$} & \colhead{$M_{\rm{ej, ext}}$} & \colhead{$f_{\rm{disk}}$}\\
\colhead{--} & \colhead{[$M_\odot$]} & \colhead{--} & \colhead{[$M_\odot$]} & \colhead{--} & \colhead{$ \left[ \frac{k_{\rm{B}}}{\rm{baryon}} \right] $} \vrule & \colhead{[$c$]} & \colhead{$ \left[ \frac{k_{\rm{B}}}{\rm{baryon}} \right] $} & \colhead{--} & \colhead{--} & \colhead{--} & \colhead{[$M_\odot$]} & \colhead{[$M_\odot$]} & \colhead{--}}
\vspace{5mm}
\startdata
\textbf{1} & \textbf{2.58} & \textbf{0.00} & \textbf{0.12} & \textbf{0.10} & \textbf{4.00} & \textbf{0.066} & \textbf{17} & \textbf{0.16} & \textbf{0.14} & \textbf{0.06} & \textbf{1.1e-03} & \textbf{8.4e-02} & \textbf{0.70} \\
\textbf{\textit{2}} & \textbf{\textit{2.58}} & \textbf{\textit{0.69}} & \textbf{\textit{0.12}} & \textbf{\textit{0.10}} & \textbf{\textit{4.00}} & \textbf{\textit{0.075}} & \textbf{\textit{17}} & \textbf{\textit{0.25}} & \textbf{\textit{0.08}} & \textbf{\textit{0.00}} & \textbf{\textit{1.9e-03}} & \textbf{\textit{9.1e-02}} & \textbf{\textit{0.75}} \\
\textbf{3} & \textbf{3.00} & \textbf{0.80} & \textbf{0.0002} & \textbf{0.50} & \textbf{8.00} & \textbf{0.064} & \textbf{89} & \textbf{0.50} & \textbf{0.00} & \textbf{0.00} & \textbf{2.8e-07} & \textbf{3.2e-07} & \textbf{0.00} \\
\textbf{4} & \textbf{3.00} & \textbf{0.80} & \textbf{0.003} & \textbf{0.10} & \textbf{8.00} & \textbf{0.063} & \textbf{29} & \textbf{0.18} & \textbf{0.17} & \textbf{0.08} & \textbf{7.6e-05} & \textbf{2.4e-03} & \textbf{0.79} \\
\textbf{5} & \textbf{3.00} & \textbf{0.80} & \textbf{0.0016} & \textbf{0.50} & \textbf{8.00} & \textbf{0.069} & \textbf{31} & \textbf{0.45} & \textbf{0.00} & \textbf{0.00} & \textbf{2.9e-05} & \textbf{1.0e-03} & \textbf{0.65} \\
\textbf{6} & \textbf{3.00} & \textbf{0.80} & \textbf{0.02} & \textbf{0.10} & \textbf{8.00} & \textbf{0.067} & \textbf{24} & \textbf{0.22} & \textbf{0.09} & \textbf{0.02} & \textbf{2.8e-04} & \textbf{7.5e-03} & \textbf{0.38} \\
\textbf{\textit{7}} & \textbf{\textit{3.00}} & \textbf{\textit{0.80}} & \textbf{\textit{0.02}} & \textbf{\textit{0.50}} & \textbf{\textit{8.00}} & \textbf{\textit{0.067}} & \textbf{\textit{18}} & \textbf{\textit{0.38}} & \textbf{\textit{0.00}} & \textbf{\textit{0.00}} & \textbf{\textit{6.6e-04}} & \textbf{\textit{1.8e-03}} & \textbf{\textit{0.09}} \\
8  & 3.90 & 0.64 & 0.289 & 0.34 & 6.56 & 0.101 & 13 & 0.36 & 0.00 & 0.00 & 1.5e-02 & 2.3e-01 & \hspace{4mm} 0.79 (*) \\
\textbf{\textit{9}} & \textbf{\textit{5.00}} & \textbf{\textit{0.90}} & \textbf{\textit{0.42}} & \textbf{\textit{0.10}} & \textbf{\textit{4.00}} & \textbf{\textit{0.083}} & \textbf{\textit{19}} & \textbf{\textit{0.32}} & \textbf{\textit{0.02}} & \textbf{\textit{0.00}} & \textbf{\textit{1.3e-02}} & \textbf{\textit{3.5e-01}} & \textbf{\textit{0.82}} \\
10  & 5.25 & 0.61 & 0.183 & 0.39 & 7.57 & 0.069 & 19 & 0.32 & 0.00 & 0.00 & 2.2e-03 & 1.4e-02 & \hspace{4mm} 0.08 (\textdagger) \\
11  & 5.59 & 0.61 & 0.231 & 0.37 & 6.96 & 0.068 & 20 & 0.34 & 0.00 & 0.00 & 2.0e-03 & 3.0e-02 & 0.13 \\
12  & 5.68 & 0.84 & 0.196 & 0.44 & 4.65 & 0.082 & 26 & 0.36 & 0.00 & 0.00 & 1.2e-03 & 1.1e-01 & \hspace{4mm} 0.54 (*) \\
13  & 5.96 & 0.79 & 0.131 & 0.27 & 6.73 & 0.072 & 20 & 0.32 & 0.00 & 0.00 & 4.6e-03 & 6.4e-02 & 0.49 \\
\textbf{\textit{14}} & \textbf{\textit{6.00}} & \textbf{\textit{0.75}} & \textbf{\textit{0.143}} & \textbf{\textit{0.15}} & \textbf{\textit{8.00}} & \textbf{\textit{0.070}} & \textbf{\textit{22}} & \textbf{\textit{0.30}} & \textbf{\textit{0.00}} & \textbf{\textit{0.00}} & \textbf{\textit{3.2e-03}} & \textbf{\textit{6.5e-02}} & \textbf{\textit{0.45}} \\
15  & 6.25 & 0.82 & 0.199 & 0.31 & 4.29 & 0.077 & 25 & 0.36 & 0.00 & 0.00 & 2.4e-03 & 1.5e-01 & \hspace{4mm} 0.74 (*\textdagger) \\
16  & 6.37 & 0.78 & 0.133 & 0.48 & 6.02 & 0.080 & 29 & 0.38 & 0.00 & 0.00 & 5.8e-04 & 6.7e-02 & \hspace{4mm} 0.51 (\textdagger) \\
\textbf{\textit{17}} & \textbf{\textit{7.00}} & \textbf{\textit{0.90}} & \textbf{\textit{0.25}} & \textbf{\textit{0.10}} & \textbf{\textit{4.00}} & \textbf{\textit{0.083}} & \textbf{\textit{27}} & \textbf{\textit{0.34}} & \textbf{\textit{0.00}} & \textbf{\textit{0.00}} & \textbf{\textit{3.9e-03}} & \textbf{\textit{2.0e-01}} & \textbf{\textit{0.79}} \\
18  & 8.13 & 0.71 & 0.044 & 0.35 & 5.07 & 0.075 & 43 & 0.34 & 0.00 & 0.00 & 2.3e-04 & 3.0e-03 & 0.07 \\
19  & 8.33 & 0.66 & 0.057 & 0.43 & 7.43 & 0.057 & 37 & 0.39 & 0.00 & 0.00 & 2.5e-04 & 2.7e-02 & 0.47 \\
20  & 8.48 & 0.77 & 0.114 & 0.34 & 6.27 & 0.071 & 30 & 0.33 & 0.00 & 0.00 & 1.3e-03 & 7.3e-02 & 0.64 \\
21  & 9.08 & 0.77 & 0.072 & 0.33 & 6.40 & 0.072 & 36 & 0.32 & 0.00 & 0.00 & 7.1e-04 & 4.8e-02 & 0.67 \\
22  & 9.38 & 0.79 & 0.064 & 0.41 & 5.22 & 0.066 & 45 & 0.37 & 0.00 & 0.00 & 2.7e-04 & 2.7e-02 & 0.43 \\
23  & 9.68 & 0.74 & 0.109 & 0.42 & 5.66 & 0.080 & 43 & 0.39 & 0.00 & 0.00 & 3.8e-04 & 4.8e-02 & 0.44 \\
24  & 9.85 & 0.68 & 0.098 & 0.49 & 4.05 & 0.074 & 52 & 0.44 & 0.00 & 0.00 & 2.2e-04 & 3.5e-02 & 0.35 \\
\textbf{\textit{25}} & \textbf{\textit{10.00}} & \textbf{\textit{0.80}} & \textbf{\textit{0.082}} & \textbf{\textit{0.20}} & \textbf{\textit{8.00}} & \textbf{\textit{0.089}} & \textbf{\textit{39}} & \textbf{\textit{0.27}} & \textbf{\textit{0.03}} & \textbf{\textit{0.00}} & \textbf{\textit{8.6e-04}} & \textbf{\textit{6.5e-02}} & \hspace{4mm} \textbf{\textit{0.80}} (*) \\
\hline
Mean & -- & -- & -- & -- & -- & 0.074 & 31 & 0.33 & 0.02 & 0.01 & 2.2e-03 & 7.1e-02 & 0.50 \\
No outliers & -- & -- & -- & -- & -- & 0.072 & 29 & 0.33 & 0.02 & 0.01 & 1.4e-03 & 5.6e-02 & 0.48 \\
Median & -- & -- & -- & -- & -- & 0.072 & 27 & 0.34 & 0.00 & 0.00 & 8.6e-04 & 4.8e-02 & 0.51 \\
No outliers & -- & -- & -- & -- & -- & 0.071 & 26 & 0.34 & 0.00 & 0.00 & 8.9e-04 & 4.8e-02 & 0.50 \\
\enddata
\label{tbl:ejecta}
\caption{Selected ejecta parameters of interest for input parameters given on the left-hand side of the table. Simulation input parameters are the same as in Table~\ref{tbl:disks}. Outputs in brackets are mass-averaged quantities, specifically the radial velocity $v_{\rm{r}}$, simulation-end entropy $s$, and the final electron fraction $Y_{\rm{e}}$. Non-mass-averaged properties include the lanthanide and actinide mass fractions $X_{\rm{La}}, X_{\rm{Ac}}$, ejected mass at simulation end $M_{\rm{ej}}$, extrapolated ejected mass $M_{\rm{ej, ext}}$, and fraction of the disk expelled as ejecta $f_{\rm{disk}}$. The same stylistic (bold and italicized text) and symbolic descriptors apply as in Table~\ref{tbl:disks}.}
\end{deluxetable*}

We present our library of 25 GR$\nu$RMHD black hole accretion disks in Tables~\ref{tbl:disks} and \ref{tbl:ejecta}. The library consists of the 11 initial disks plus 14 GP-placed disks (down from ${\sim} 25$ GP-placed disks after vetting for disk stability, simulation artifacts, etc.). The left side of each table indicates the simulation input parameters consisting of the black hole mass $M_{\rm{bh}}$ and spin $a_{\rm{bh}}$ and the accretion disk's initial mass $M_{\rm{disk}}$, electron fraction $Y_{\rm{e}}$, and entropy $s$. Each disk is numbered for convenient reference. All quantities in brackets $\langle ... \rangle$ represent mass-averaged quantities. The right side of each table presents a selection of disk (Table~\ref{tbl:disks}) and ejecta (Table~\ref{tbl:ejecta}) properties of interest, with some calculated at the time of peak accretion onto the black hole and the remainder calculated at the end of the simulation at $t = 10^4M_{\rm{bh}}$. In Table~\ref{tbl:disks}, we provide the peak accretion rate $\dot{M}_{\rm{peak}}$, mass-weighted temperature at peak accretion rate $\langle T\rangle$, mass-weighted density at peak accretion $\langle \rho \rangle$, the maximum Blandford-Znajek jet power $P_{\rm{jet, max}}$ \citep{BZ77, MT82, 2010ApJ...711...50T}, maximum neutrino luminosity at the time of peak accretion $L_{\nu, {\rm{max}}}$, the dimensionless Shakura--Sunyaev effective viscosity parameter $\alpha$ at the time of peak accretion and radius of peak pressure, and the extrapolated disk lifetime $\tau$. In Table~\ref{tbl:ejecta}, we provide the radial velocity $v_r$, electron fraction $Y_{\rm{e}}$, entropy $s$, lanthanide mass fraction $X_{\rm{La}}$, actinide mass fraction $X_{\rm{Ac}}$, ejected mass by simulation end $M_{\rm{ej}}$, extrapolated ejected mass by disk lifetime ($\tau$) end $M_{\rm{ej, ext}}$, and the fraction of the total disk mass ejected $f_{\rm{disk}}$.

The formatting of the entries in each table includes bold and italicized text and symbolic markers to delineate specific simulations. The majority of the disks in Tables~\ref{tbl:disks} and \ref{tbl:ejecta} are being presented in the literature for the first time. However, Disks 2, 7, 9, 14, 17, and 25 (or modified versions of them) have previously been presented in the literature. Namely, Disk 2 is presented in \cite{2019PhRvD.100b3008M}, in the same form as in this work, to study the blue kilonova associated with GW170817. \cite{2024ApJ...962...79S} extend the Disk 2 simulation roughly ten-fold out to 1.2 seconds post-merger with associated late-time nucleosynthesis calculations. \cite{2024ApJ...964..111L} and \cite{2025ApJ...985L...9L} use the Disk 2 simulation as a baseline for comparison of magnetic field and neutrino fast flavor transformation effects on nucleosynthesis, respectively. Disk 7 is presented in \cite{2020ApJ...902...66M}, in the same form as in this work, examining nucleosynthesis in a collapsar environment. Lastly, Disks 9, 14, 17, and 25 are examined in \cite{2023ApJ...945L..13C} within the context of black hole-neutron star post-merger disk nucleosynthesis. These previously presented disks and their associated parameters are italicized in Table~\ref{tbl:disks} to distinguish them from the newly presented simulations in this work. Disks which are used for the initial GP training set, prior to any simulation placement, are marked in boldface font. Some simulations terminated prior to reaching $t = 10^4M_{\rm{bh}}$ and thus may represent a different disk state (possibly with unsaturated magnetorotational instability) than those which ran to completion; these simulations are marked by an asterisk (*). Finally, the last few simulations we ran (in order of simulation placement, not table position) were \textit{fully three-dimensional}, but their outputs required downsampling to axisymmetry (two dimensions, $r$ and $\theta$) due to data storage constraints; these are identified with a dagger (\textdagger).

The majority of the columns in Table~\ref{tbl:disks} report values at the time of peak disk accretion onto the black hole. For $\dot{M}_{\rm{peak}}$, the peak accretion rate, the largest value is achieved by the heaviest disk (Disk 9), though $\dot{M}_{\rm{peak}}$ does not scale one-to-one with disk mass in general. The mass-weighted tracer temperature at peak accretion $\langle T \rangle$ averages above 10 GK (above which we can safely assume the tracer is in NSE); while some disks report lower temperatures at the time of peak accretion, all tracers used for nucleosynthesis were confirmed to begin nucleosynthesis above 10 GK. For $\langle \rho \rangle$, we find that, at the time of peak accretion, the density loosely rises with the disk mass; this is in contrast to the maximum density, which scales nearly one-to-one with disk mass. The maximum jet power at accretion peak $P_{\rm{jet, max}}$ scales nearly linearly with the disk mass, though given the use of the Blandford--Znajek mechanism to calculate the jet's power, it is surprising that Disks 8 and 9 have quite different $\dot{M}_{\rm{peak}}$ but very comparable maximum jet powers. Our $\alpha$ viscosity values are similar to those reported by \cite{2021ApJ...921...94D} given that the radius of peak pressure is typically located around $\mathcal{O}(10)$ km; notably, the non-spinning case exhibits a higher-than-usual $\alpha$, implying the radius of peak pressure was much farther in, closer to the horizon. 

For the ejecta values in Table~\ref{tbl:ejecta}, the most notable outlier across the board is Disk 3, with its very low mass and observationally insignificant mass ejection. The mass-averaged radial velocity $\langle v_r \rangle$ is consistent across the majority of disks; of the disks which did not run to completion, only Disks 8 and 25 show appreciably higher ejecta velocities. The joint impact of entropy $s$, electron fraction $Y_{\rm{e}}$, and various nucleosynthesis yields (e.g., $X_{\rm{La}}$ and $X_{\rm{Ac}}$) is discussed in detail in Section~\ref{sec:discussion}. Here, we wish to emphasize that the values for $X_{\rm{La}}$ and $X_{\rm{Ac}}$ presented here assume no cuts on $v_r$ and $\theta$ prior to a random tracer draw for each of 15 $Y_{\rm{e}}$ bins. While these values do not fully represent the nucleosynthesis associated with the tracer population from each disk, we present them here for completeness and point readers to Figures~\ref{fig:vrlt_hist}, \ref{fig:vrgt_hist}, \ref{fig:thetalt_hist}, and \ref{fig:thetagt_hist} for histogram plots of tracer mass as a function of $(Y_e, \theta)$. The final three columns represent $M_{\rm{ej}}$ at simulation end (i.e., the amount of mass assumed to escape for tracers which meet the Bernoulli criterion), $M_{\rm{ej, ext}}$, the extrapolated ejected mass per the methodology described in Section~\ref{sec:mej_ext}, and $f_{\rm{disk}}$, the fraction of the disk mass which has been ejected rather than accreted.

We examine the relationship between peak jet power and the accretion disk mass, radial velocity $v_r$ distributions, mass ejection trends as a function of disk mass, and relationship between nucleosynthesis and $Y_{\rm{e}}$ in more detail in Sections~\ref{sec:peak_jet_power}--\ref{sec:nucleosynthesis} below. 

\subsection{Peak Jet Power vs. Initial Disk Mass}
\label{sec:peak_jet_power}

\begin{figure}
    \centering
    \includegraphics[width=\linewidth]{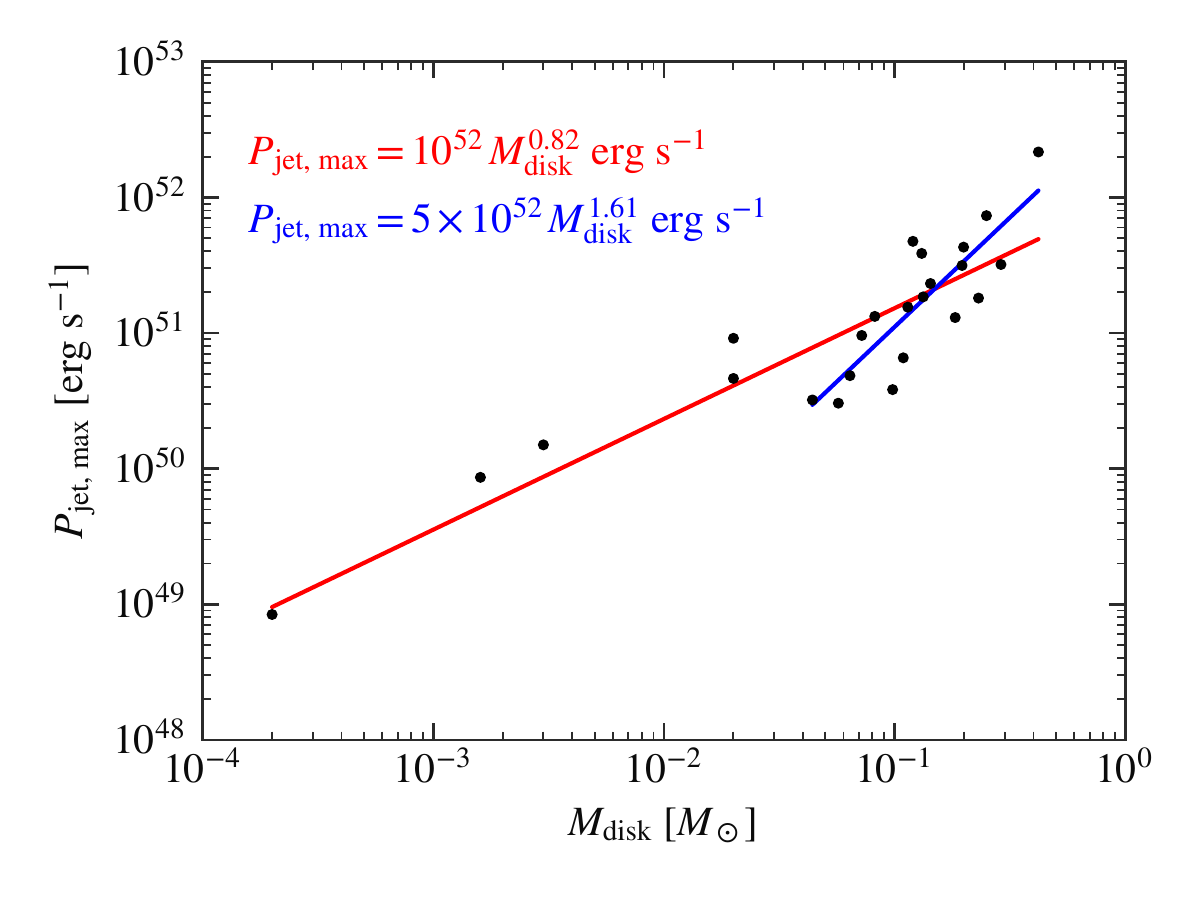}
    \caption{Peak Blandford-Znajek jet power as a function of disk mass, plotted in log-log coordinates. In the context of kilonova detections with associated gamma-ray bursts, this relation can be used to place an upper limit on the accretion disk mass and, by extension, the post-merger ejecta mass.}
    \label{fig:jet_power}
\end{figure}

We examine the relationship between the accretion disk mass and the peak jet power in Figure~\ref{fig:jet_power}. While our simulations do not model a jet directly, we calculate the jet power through the Blandford--Znajek process \citep{BZ77, MT82, 2010ApJ...711...50T}. In focusing specifically on the peak jet power, our goal is to connect the gamma-ray burst (GRB) observations to the circumremnant accretion disk assuming a compact object or collapsar source. This relation is not meant to serve as a definitive classifier of disk properties from a kilonova source event; instead, it provides an approximate upper-bound estimate on the disk mass, and thus the post-merger ejecta, associated with a given kilonova detection. We find the following relation between peak jet power and accretion disk mass: 
\begin{equation}
P_{\rm{jet, max}} = 10^{52} M_{\rm{disk}}^{0.82} \text{ erg s}^{-1},
\end{equation}
with a coefficient of determination $R^2 = 0.804$ assuming $M_{\rm{disk}}$ in units of $M_\odot$. The fit shifts to a more positive slope for $\log M_{\rm{disk}} > -1.5$ with a coefficient of $5 \times 10^{52}$ and an exponent of $1.61$.

\subsection{Disk Ejecta Velocity Distribution}
\label{sec:vr}

\begin{figure}[ht!]
    \centering
    \includegraphics[width=\linewidth]{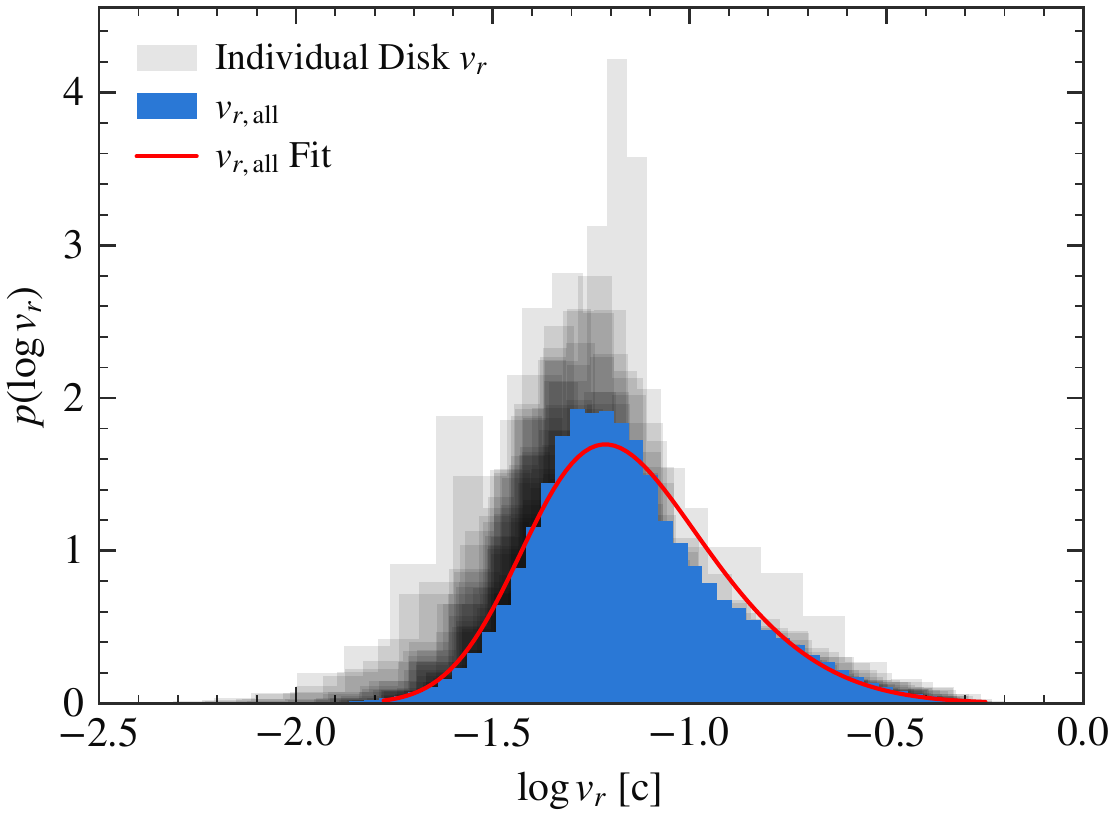}
    \caption{Mass-weighted tracer radial velocity $v_r$ distributions from the individual disks (gray) and the concatenation of all disks' tracers together (blue). The red line is a skew-normal distribution fit to the concatenated blue histogram, with parameters $\alpha = 2.01$, $\mu = -1.4$, and $\sigma$ = 0.35. }
    \label{fig:vrfit}
\end{figure}

Nearly all disks have an end-time mass-weighted radial velocity $\langle v_r \rangle$ that lies within a consistent range despite the black holes and disks having appreciably different physical parameters.
We investigate this phenomenon in Figure~\ref{fig:vrfit}. Initially, we plot the $v_r$ histograms for each individual disk, shown by the translucent gray histograms in the background. We concatenate the $v_r$ and mass weights for all tracers across all simulations to generate the blue, mass-weighted $v_r$ histogram representative of the average $v_r$ distribution across all disks.
Although this distribution is nearly Gaussian, there is a longer tail tending toward higher-velocity ejecta which can have substantial impact on the thermodynamic and spectroscopic evolution of the ejecta.
To account for this longer tail, we fit the disk-averaged $v_r$ distribution with a skew normal distribution $p(x, \alpha, \mu, \sigma) \equiv p(y, \alpha)/\sigma$ such that our $\log v_r$ distribution is characterized by
\begin{equation}
    p(\log v_r) \approx \frac{2}{\sigma} \phi(y) \Phi(\alpha y)
\end{equation}
where $y = (x - \mu)/\sigma$ and $\phi(y)$ and $\Phi(\alpha y)$ are the standard normal probability density and cumulative distribution functions, respectively.
The fit in Figure~\ref{fig:vrfit} is characterized by a skew normal distribution with skewness parameter $\alpha = 2.01$ with fixed location $\mu = -1.4$ and scale $\sigma = 0.35$.

\begin{figure}[ht!]
    \centering
    \includegraphics[width=\linewidth]{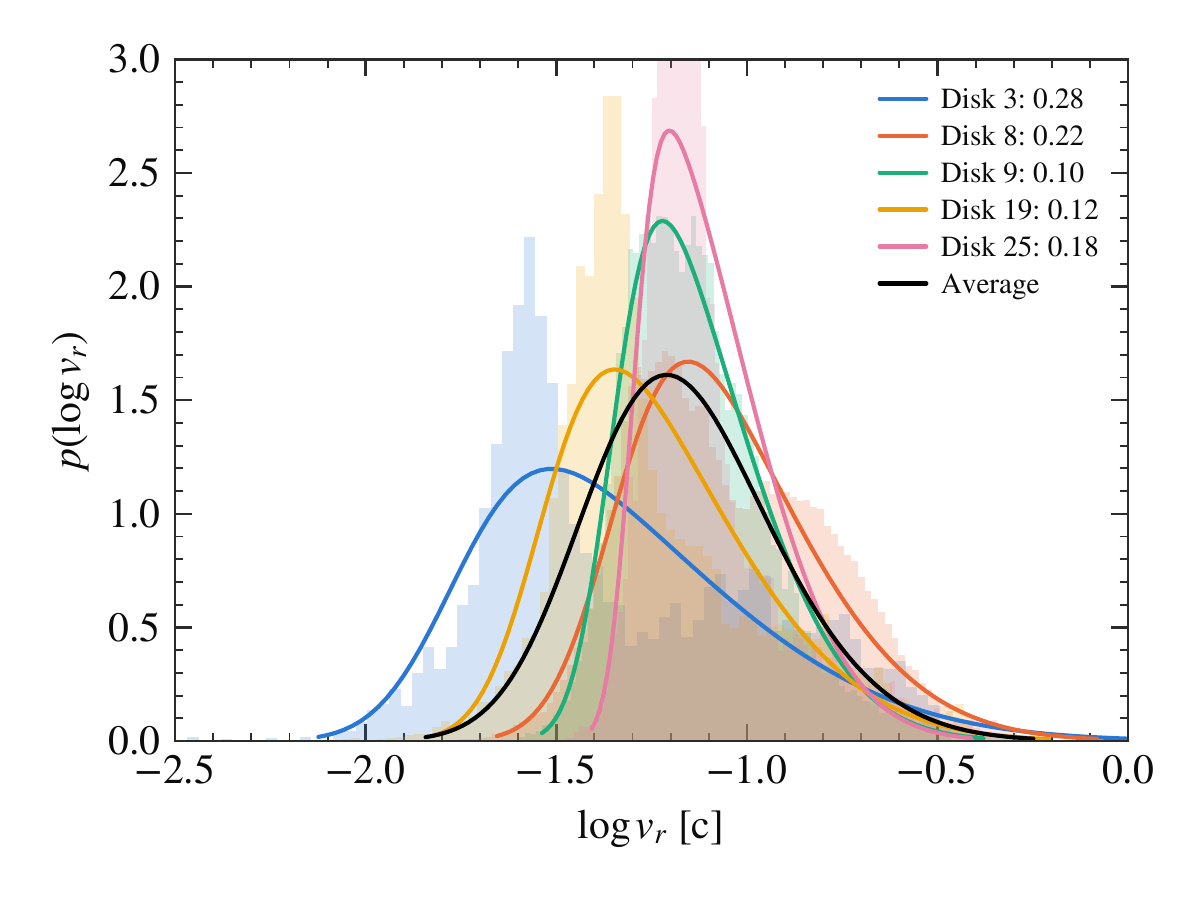}
    \caption{Velocity distribution skew-normal fits and histograms for disks which deviate appreciably (JS divergence $> 0.1$) from the average fit presented in Figure~\ref{fig:vrfit}. The legend indicates each disk and its JS divergence. Disk 3 is the lightest by an order of magnitude, while Disks 8 and 9 are the two heaviest. Disks 8 and 25 terminated prior to $t = 10^4M_{\rm{bh}}$.}
    \label{fig:outliers}
\end{figure}

To assess how well the $v_r$ fit describes the individual disks' $v_r$ distributions, we employ a Jensen-Shannon (JS) divergence analysis of the velocity distributions. In essence, the JS divergence compares the similarity between two probability distributions, evaluating to zero for equivalent distributions.
We find that most disks are well represented by the fit to $v_r$, with JS divergences below 0.10 save for a few outliers, namely Disks 3, 8, 9, 19, and 25. 
The skew normal fits to the outliers' $v_r$ distributions are shown in Figure~\ref{fig:outliers}.

\subsection{Disk Ejecta Mass vs. Initial Disk Mass}
\label{sec:meje}

\begin{deluxetable}{c|cc}[hb!]
\centering
\tablewidth{0pt}
\tablecolumns{3}
\tablehead{
\colhead{Quantity} & \colhead{$A \ [M_\odot]$} & \colhead{$p$}
}
\startdata
$M_{\rm{ej}}$ & $1.10^{+0.22}_{-0.19} \times 10^{-3}$ & $1.13 \pm 0.10$ \\
$M_{\rm{ej, ext}}$ & $4.24^{+1.10}_{-0.88} \times 10^{-2}$ & $1.44 \pm 0.13$ \\
$M_{\rm{ej}}$ (reduced) & $9.82^{+1.80}_{-1.50} \times 10^{-4}$ & $0.83 \pm 0.12$ \\
$M_{\rm{ej, ext}}$ (reduced) & $3.86^{+0.82}_{-0.67} \times 10^{-2}$ & $0.98 \pm 0.14$ 
\enddata
\label{tbl:mej_fits}
\caption{Fitting coefficients for Equation~\ref{eq:meje} which predict the ejected mass at simulation end ($M_{\rm{ej}}$, measured) and at the end of the disk's lifetime ($M_{\rm{ej, ext}}$, extrapolated). The rows marked as ``(reduced)" indicate fits which omit the outlier disks shown in Figure~\ref{fig:outliers}.}
\end{deluxetable}

\begin{figure}[ht!]
    \centering
    \includegraphics[width=\linewidth]{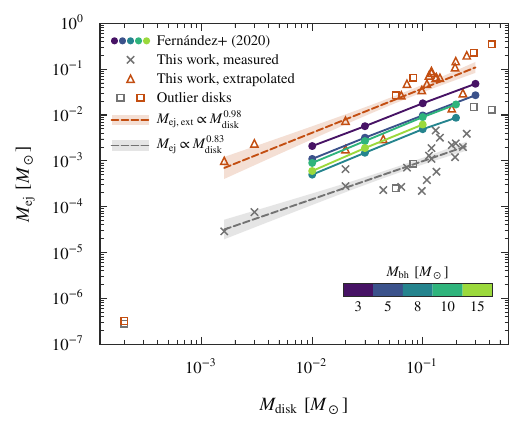}
    \caption{Ejected mass at simulation end (gray points) and ejecta mass extrapolated to the end of the disk's lifetime (orange points) as a function of disk mass. The lines and shaded regions correspond to power-law fits and their associated $1\sigma$ confidence intervals. Square markers delineate the outlier disks from Figure~\ref{fig:outliers} which are \emph{excluded from these fits}. We include the disk ejecta mass predictions following black hole-neutron star mergers from \cite{2020MNRAS.497.3221F} which span $0.92 \leq p \leq 1$, varying with remnant black hole mass.}
    \label{fig:meje}
\end{figure}

We further investigate the relationship between ejected mass and initial disk mass given the importance of ejecta mass in kilonova modeling. We report both the ejected mass at simulation end $M_{\rm{ej}}$ and the extrapolated ejected mass $M_{\rm{ej, ext}}$ that is predicted at the end of the disk's extrapolated lifetime $\tau_{\rm{ext}}$. We perform a separate power-law fit for each type of ejecta (measured and extrapolated) assuming that 
\begin{equation}
M_{\rm{ej}} = A \times \left(\frac{M_{\rm{disk}}}{0.1 M_\odot}\right)^p
\label{eq:meje}
\end{equation}
for fitting coefficients $A$, in units of $M_\odot$, and $p$. The fitting coefficients for each case are reported in Table~\ref{tbl:mej_fits} and the results for a subset of the fits are plotted in Figure~\ref{fig:meje}. Here, $M_{\rm{disk}}$ is the initial disk mass. In the figure, the gray points represent the $M_{\rm{ej}}$ simulation data as a function of $M_{\rm{disk}}$, with the line and shaded region representing the fit and associated $1\sigma$ confidence interval; the same is shown for $M_{\rm{ej, ext}}$ in orange. Additionally, we include the ejecta mass predictions from the disks reported in \cite{2020MNRAS.497.3221F} as a reference, coloring their fits as a function of their remnant black hole masses. 

In terms of other disk mass-dependent quantities, we notice a weak anticorrelation between specific entropy $\langle s \rangle$ and initial disk mass $M_{\rm{disk}}$, noting that lighter disks tend to have larger specific entropy at simulation end than heavier ones. Namely, the lightest Disk 3 has $\langle s \rangle \sim 90$, while the two heaviest Disks 8 and 9 have $\langle s \rangle \sim$ 13 and 19, respectively. If we consider initial disk specific entropy $s$ and examine the final-to-initial specific entropy ratio $\langle s \rangle / s$, we find that Disks 24 and 3 have the largest ratios, noting again that Disk 3 is the most significant outlier in terms of the velocity distribution. 

\subsection{Disk Ejecta Nucleosynthesis}
\label{sec:nucleosynthesis}

Figures~\ref{fig:vrlt}, \ref{fig:vrgt}, \ref{fig:thetalt}, and \ref{fig:thetagt} show the nucleosynthesis yields as mass fractions $X$ as a function of mass number $A$ for the four different $v_r$ and $\theta$ combinations described in Section~\ref{sec:prism}: $v_r < 0.08c$, $v_r \geq 0.08c$, $\theta < 30^\circ$, and $\theta \geq 30^\circ$, respectively.
Each subplot has two gray vertical bands indicating the regions where lanthanides ($139 \lesssim A \lesssim 175$) and actinides ($227 \lesssim A \lesssim 262$) reside as well as gray dashed lines indicating the first ($A \sim 80$), second ($A \sim 130$), and third ($A \sim 195$) $r$-process peaks \citep{2021RvMP...93a5002C}. The subplots are colored as a function of initial disk $Y_{\rm{e}}$ using the typically assumed ansatz that $Y_e < 0.25$ produces a lanthanide-rich ejecta leading to a ``red" kilonova, while $Y_e \geq 0.25$ produces a lanthanide-poor ejecta and a ``blue" kilonova. 
This color scheme is chosen in part to highlight that this convenient split is not all-encompassing, as a few of the disks with $0.1 \leq Y_e \leq 0.2$ fail to produce lanthanides.  The associated mass distributions for each cut as a function of $Y_{\rm{e}}$ and $\theta$ are shown in Figures~\ref{fig:vrlt_hist}, \ref{fig:vrgt_hist}, \ref{fig:thetalt_hist}, and \ref{fig:thetagt_hist}.

\section{Discussion}
\label{sec:discussion}

We begin with a comparison of our results to those from previous work mentioned in Section~\ref{sec:intro}. We find good agreement between our characteristic value of $\langle v_r/c \rangle \sim 0.07$ and $v_r/c \sim 0.067$ reported in \cite{2013MNRAS.435..502F}. Our velocities are slower than the $\mathcal{O}(0.1c)$ velocities from the three-dimensional, GRMHD simulation presented in \cite{2019MNRAS.482.3373F} and the poloidal, strong magnetic field simulation in \cite{2019MNRAS.490.4811C}, but faster than the former's two-dimensional simulations. The simulations in our study, which have been reduced to two dimensions, exhibit consistently faster velocities than the two-dimensional runs of \cite{2019MNRAS.482.3373F}; our results loosely suggest that reducing simulation data to two dimensions after simulating in three does not bias the velocity distribution. Our radial velocity trends have a roughly $10\%$ spread at our extraction surface of $r=250M_{\rm{bh}}$, corresponding to 1--4$ \times10^3$ km. They are 40--110\% larger than the mean velocities evaluated at $r = 10^4$ km in \cite{2015MNRAS.448..541J}, but closely aligned with the characteristic value of $20,000$ km/s (${\sim}0.067c)$ at the same radius obtained by \cite{2013MNRAS.435..502F}.

\begin{figure*}
\includegraphics[width=\linewidth]{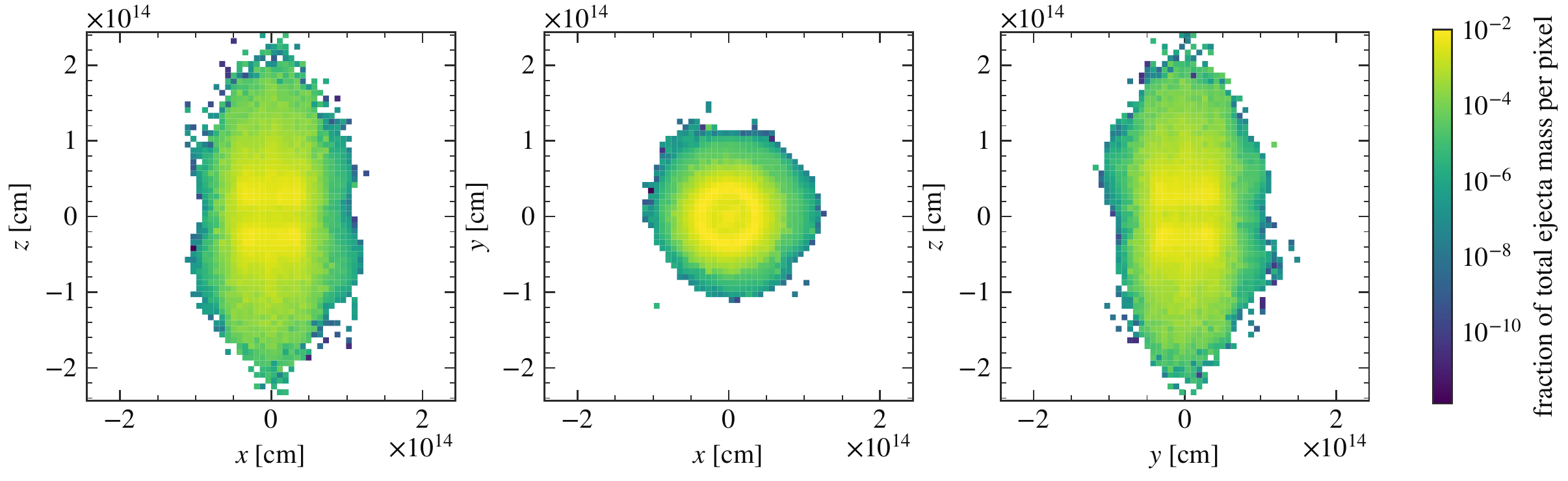}
\caption{Cross-sections in the x-z, x-y, and y-z planes of the disk ejecta morphology extrapolated out to three hours ($10,800$ seconds) post-merger when averaged across the simulation library. The colorbar represents the fraction of mass in each pixel. The ejecta shape is distinctly double-lobed along the jet-axis, favoring a non-spherical geometry for the post-merger ejecta component in kilonova simulations.}
\label{fig:morph}
\end{figure*}

Our extrapolated ejecta mass from the disk is substantially larger ($M_{\rm{ej, ext}} \sim 5 \times 10^{-2} M_\odot$) for shorter average disk lifetimes ($\tau_{\rm{ext}} \sim 2$ s) than the ejected mass from the few-second simulations, suggesting that we start extrapolating the disk lifetime before the disk has settled into a stable outflow state. Figure~7 of \cite{2019MNRAS.482.3373F} shows their GRMHD outflow rate reaching a stable state described by a power-law around ${\sim}1$ second, which none of our simulations reach. In spite of this more significant mass ejection, our extrapolated values for the ejected disk fraction average around 50\%, which is only a modest 10--20\% higher than the fractions presented in the previous works described in Section~\ref{sec:intro}. Specifically, in Figure~\ref{fig:meje}, we compare our results to the disk ejecta from the black hole-neutron star merger simulations presented in \cite{2020MNRAS.497.3221F}. We find that our measured ejecta fit deviates from the trends that they found, but our extrapolated ejecta power-law exponent is consistent with their fits for all remnant $M_{\rm{bh}}$. Our coefficient $p=0.97$, derived from a variety of black hole and disk masses, lands within their fits which span $0.92 \leq p \leq 1$ depending on the remnant black hole mass, supporting their finding that heavier disks eject a smaller fraction of their mass. However, our extrapolation strategy is imperfect, and this agreement should be cautiously considered. Compared to the recent end-to-end, first-principles study performed by \cite{2026arXiv260827555K}, only eight of our disks expel enough post-merger ejecta (assuming an arbitrary dynamical ejecta contribution of $\mathcal{O}(10^{-2}M_\odot)$) to match their total ejected mass of $0.08 M_\odot$. Future studies examining a subset of these disks at longer simulation times will better characterize the ejected disk fraction using more accurate disk lifetimes.

The ejecta morphologies at the end of our simulations are still dynamic; while accretion and mass ejection rates are more stable relative to those at $t \lesssim 3000M_{\rm{bh}}$, they are still prone to change at longer simulation times prior to settling into a stable state (see again the outflow rate in Figure~7 of \cite{2019MNRAS.482.3373F}). With this caveat in mind, we nevertheless extrapolate tracer information from our extraction surfaces at $r = 250M_{\rm{bh}}$ to explore the possible disk ejecta morphology diversity from our simulation library. We assume that the tracer trajectory at the extraction surface remains constant over the duration of extrapolation as well; in other words, we assume that the tracer's Cartesian velocity remains constant and that the tracer is subject to no external forces thereafter. While we analyze the morphology for all disks individually, we only report the disk library-averaged morphology in this work. In Figure~\ref{fig:morph}, we present cross-sections in the x-z, x-y, and y-z planes for a disk-averaged ejecta morphology extrapolated to three hours ($10,800$ seconds) post-merger, roughly around the onset of kilonova emission. Each disk is equally weighted during averaging to prioritize understanding the average shape the ejecta takes across all disks. A similar analysis using mass-weighting biased the shape in favor of more massive disks; the equal-disk weighting we employ here, in turn, favors disks with more tracers (in some cases, tracers differ by an order of magnitude). In general, we interpret the results of Figure~\ref{fig:morph} and similar analyses to motivate continued use of the post-merger ejecta representation as a double-lobed (``peanut-shaped") morphology along the jet axis \citep[e.g.][]{2021ApJ...910..116K, 2021ApJ...918...10W, 2022PhRvR...4a3046R}.

We find a wide range of final $Y_{\rm{e}}$ values with $0.16 \leq \langle Y_{\rm{e}} \rangle \leq 0.50$ and disk-library-averaged values that robustly support $\langle Y_{\rm{e}} \rangle \sim 0.34$, larger than the $\langle Y_{\rm{e}} \rangle \sim 0.20$ reported in \cite{2013MNRAS.435..502F}. We note, however, that this value might be influenced by the majority of newly placed disks (non-bolded entries in Table~\ref{tbl:disks}) floating around an initial disk $Y_{\rm{e}} \sim 0.38 \pm 0.06$, with six out of the 14 GP-placed disk initial conditions close to the final $\langle Y_{\rm{e}} \rangle$ values. Omitting GP-placed disks with $Y_e = 0.34 \pm 0.05$ yields a disk-library-averaged $\langle Y_e \rangle = 0.33$, supporting the robustness of the result.

Our nucleosynthesis results corroborate those of \cite{2014MNRAS.441.3444M} and \cite{2018ApJ...860...64F} in that all disks with $\langle Y_{\rm{e}} \rangle > 0.25$ produce no significant amount of lanthanides under any circumstances and potentially support the production of a rapidly evolving blue transient, though with the caveat that $Y_{\rm{e}}$ tends to shift toward lower values for smaller $\theta$ (see Figures~\ref{fig:thetalt_hist} and \ref{fig:thetagt_hist}). Only Disks 1, 4, and 6 produce meaningful amounts of actinides ($X_{\rm{Ac}} > 10^{-4}$), though it is unclear whether all of them are observationally significant. Disks 4 and 6 show additionally suppressed actinide production for equatorial tracers ($\theta \geq 30^\circ$). While Disks 1, 2, 4, 6, and 9 are all around light black holes ($M_{\rm{bh}} \leq 5 M_\odot$) and have the same low initial $Y_{e} = 0.1$, we find that actinide production is innately tied to the black hole spin and initial disk entropy, with Disks 1, 4 and 6 producing actinides in some capacity and starting with no spin (Disk 1) or high initial entropy $s = 8$ k$_B$/baryon. Conversely, Disks 2 and 9 start with low initial entropy $s = 4$ k$_B$/baryon and never produce substantive amounts of actinides. The same lack of actinide production holds for Disk 17 which starts with the same low entropy and low initial $Y_{\rm{e}}$, but is formed around a heavier black hole with $M_{\rm{bh}} = 7M_\odot$. These trends can also be observed in the $Y_{\rm{e}}$ histograms such as in Figure~\ref{fig:thetagt_hist}, where the $Y_{\rm{e}}$ distributions for Disks 1, 4, and 6 have tails that terminate at lower $Y_{\rm{e}}$ compared to the distributions for Disks 2, 9 and 17. For disks which produce third-peak $r$-process elements, we find that lighter black holes do produce more third-peak elements, though Disk 25 around the heaviest black hole does produce a sub-percent mass fraction of third-peak elements. In general, our findings largely support the inverse relationship between black hole mass and third-peak strength presented in \cite{2015MNRAS.448..541J}, especially when considering observationally relevant quantities of element production. 

The $v_r$ fit outliers are an interesting case in that there is no immediately obvious pattern that may predict or distinguish which disks deviate from the fit.
Disks 8 and 25 are both only simulated out to ${\sim} 7.3 \times 10^3 M_{\rm{bh}}$, potentially indicating that, if continued for the entire duration of the simulation, their disks would relax and produce outflows more closely resembling the average $v_r$ distribution.
Yet, these two disks have very different $v_r$ distributions at the same number of light-crossing times, with Disk 8 much closer in shape and appearance to the average than the Disk 25 $v_r$ distribution's sharper peak. 
The $v_r$ distribution for Disk 9 is much more similar in both shape and width to that of Disk 25. While Disk 9 ran to the full $t = 10^4M_{\rm{bh}}$ duration, it also had by far the most massive disk $M_{\rm{disk}} = 0.42 M_{\odot}$, potentially requiring additional time compared to the other disks to relax into the average $v_r$ distribution, assuming that would indeed happen given sufficient simulation time.
In a similar vein, Disk 3 is the lightest disk in the library by an order of magnitude, with these two extrema hinting at the possibility of a connection between disk mass and $v_r$ relaxation time.
However, Disk 8 is the second-most massive disk in the library and nearly matches the average $v_r$ distribution; as such, no concrete conclusions can be drawn regarding a clear connection between disk mass and the $v_r$ distribution without further detailed study, which we leave for future work.
In general, we find that for $M_{\rm{disk}} \sim \mathcal{O}(0.1 M_\odot)$, our $v_r$ distribution fit encapsulates expected tracer behavior.

The extrapolated ejecta masses and accretion rates are calculated as described in Section~\ref{sec:mej_ext}. As mentioned in that section, our derivation of $t_{\rm{shift}}$ can potentially introduce unphysical mass ejection start times in some cases. Specifically, after accounting for $t_{\rm{shift}}$, Disks 8, 10, 12, 13, and 15 start at $-3.5$, $-0.2$, $-3.1$, $-5.8$, and $-4.5$ ms, respectively. Except for Disk 13, all of these disks terminated sooner than expected, were reduced to 2D axisymmetry, or both. As with the $v_r$ distribution outliers, it is difficult to make steadfast claims about whether these effects are directly contributing to unphysical shift times for these particular disks. Though unphysical, most of these time shifts are only pushing tracer start time back a few ms, roughly the same timescales required for the disk to start accreting. As such, the negative time shift is not expected to contribute too adversely to extrapolated masses. The same calculation can be done using $v_{\rm{max}}$ as the fastest tracer in the $v_r$ distribution of a given disk, thus placing a lower limit on $t_{\rm{shift}}$ without incurring nonphysicality.

While it is difficult to establish a clear tie between initial and final mass-averaged entropy, especially when correcting for the final-to-initial entropy ratio, one clear entropy outlier remains in Disk 3, the lowest-mass disk. Its final entropy and entropy ratio are both large; it also has the largest JS divergence when compared to the average $v_r$ distribution, with the high-velocity tail skewing the mass-averaged $v_r$ in line with the average distribution, but with a distribution peak appreciably lower than those of all other disks. This disk's accretion rate evolution does not follow a similar trend to that of the other disks, and thus its extrapolated ejecta mass is likely far off from what is expected from a fully evolved simulation, as hinted at by the sub-1\% disk ejection fraction. Previous studies using \texttt{$\nu$bhlight} \citep[e.g.,][]{2020ApJ...902...66M} have discussed the effect of the disk's initial disruption on the temperature and density behavior. As the disk disrupts and begins to accrete, temperature and density increase and, consequently, so does the leptonic depth to neutrinos. This behavior leads to an overall suppression of neutrino emission and thus a slower deleptonization process, resulting in a slower initial decrease in $Y_{\rm{e}}$. This material then slowly releptonizes toward the equilibrium $Y_{\rm{e, eq}}$ which is dependent on the tabulated neutrino emissivities and opacities. In the case of Disk 3, however, the low initial disk mass results in low densities and temperatures, leading to lower neutrino emissivity and opacity, causing the behavior to be qualitatively different. This results in Disk 3's $Y_{\rm{e, eq}}$ being much higher than that of other disks with similar initial $Y_{\rm{e}}$, as evidenced by Disk 3 having $\langle Y_e \rangle = 0.50$ as opposed to other disks with initial $Y_{e} = 0.50$ which have at least 10\% lower $\langle Y_e \rangle$. In other words, Disk 3 has too low a temperature to ignite neutrino-driven cooling, as the weak equilibrium value of $Y_{\rm{e}}$ is always large for this disk. These behaviors, largely resulting from the disk's low initial mass, point to lower-density physical effects which are not appropriately captured by the assumptions that hold for more massive disks with $M_{\rm{disk}} \sim \mathcal{O}(0.1 M_\odot)$. 

Unfortunately, the majority of the information regarding disks which exhibited instability from their initial conditions or shortly after simulation start was lost due to computing-cluster changeover during the preparation of this work. All that remains is the last batch of 7 disks, of which Disks 10, 15, and 16 successfully ran. The remaining 4 disks failed to start; while 7 disks are insufficient for a detailed statistical analysis, it is interesting to note that, for this final batch specifically, the disks which ran successfully had $M_{\rm{disk}} < 0.20 M_\odot$ and the failed disks spanned $0.28 M_\odot \leq M_{\rm{disk}} \leq 0.38 M_\odot$. The black hole spins of the successful runs were not substantially lower so as to encourage more stable simulations, nor were their entropies particularly low to encourage more stable Fishbone--Moncrief torus initialization. Regardless, the existence of Disk 9 as the most massive disk with the largest spin value in our grid implies that spin and disk mass need not affect stability within the parameter ranges explored in our grid. It is also worth noting that, despite weighting toward lower $Y_{\rm{e}}$, the average $Y_{\rm{e}}$ of the placed simulations is around $Y_e \sim 0.38$, suggesting a possible stability dependence on $Y_{\rm{e}}$ for the Fishbone-Moncrief tori generated in this study. Future understanding of disk stability as a function of initial conditions and equation of state assumptions stands to benefit from agentic AI workflows. These systems can systematically initiate grids of disk simulations and track their stability for the first $t \sim 100M_{\rm{bh}}$ based on various initial condition sweeps, enabling large-scale systematic dependency studies. Such workflows have the potential to greatly inform our understanding of the complex black hole accretion disk parameter space.

Finally, in this work we have employed the \cite{Foucart18} prescription as the criterion for disk mass selection in our model grid. While this choice is suitable for our study, which sought to sample a broad range of accretion disk scenarios, more recent prescriptions \citep[e.g.,][]{2020PhRvD.101j3002K, 2022CQGra..39a5008N} are recommended for studies exploring realistic disks specifically in the context of binary neutron star mergers.
In future studies, we will employ these updated prescriptions to also serve as priors during simulation placement to give the uncertainty-guided GP placement more direction rather than allowing it to sample entirely based on variance-guided numerics without regard for physicality.

\section{Conclusion}
\label{sec:conclusion}

We present the largest library of GR$\nu$RMHD accretion disks around stellar-mass black holes to date, representative of the conditions following neutron star merger, black hole-neutron star merger, and collapsar events. The majority of the disks' parameters were chosen through an optimized simulation placement procedure which reduces variance across objectives of interest, in our case ejecta masses across different spatial regions surrounding the disk. Our primary results convey quantities of interest in both the disk and ejecta at simulation end and connect the peak jet power to the disk mass. Additionally, we explore the radial velocity distributions $v_r$ for each disk, providing a fit to the average $v_r$ distribution and a statistical comparison of how well each disk's distribution is represented by the average through a Jensen--Shannon divergence calculation. We identify outlier disks which deviate from the average $v_r$ distribution and discuss the simulation and physics assumptions which might contribute to their deviation. We also present ejected mass values at simulation end and extrapolated mass ejection quantities, including the ejected fraction of the disk, which we find to be $\sim50\%$ (with significant caveats). We present detailed, mass-weighted nucleosynthesis yields for
r each disk given a subset of tracer properties based on tracer velocity $v_r$ and angle $\theta$. For each disk and set of tracer property subsets, we also present associated mass histograms as a function of $Y_{\rm{e}}$ and latitude angle $90^\circ - \theta$ which help intuitively explain the nucleosynthesis behavior as a function of the tracer properties. The results from this simulation library can and should be used to guide more physically informed spatial $Y_{\rm{e}}$ distributions for downstream kilonova modeling. The information presented in this work is only a summary of the plethora of data provided by these simulations; we share additional summarized tracer information at \cite{zenodo}. Due to the sheer data volume ($\sim 1$ TB per simulation) of tracer information, detailed tracer data cannot be reasonably uploaded onto an online repository. As such, detailed tracer data will be made available through contact with the corresponding author.

\begin{acknowledgements}

This work made use of the following software packages: \texttt{matplotlib} \citep{Hunter:2007}, \texttt{numpy} \citep{numpy}, \texttt{Python} \citep{python}, \texttt{scipy} \citep{2020SciPy-NMeth}, \texttt{Cython} \citep{cython:2011}, \texttt{h5py } \citep{collette_python_hdf5_2014}, and \texttt{scikit-learn} \citep{scikit-learn,sklearn_api}.  Software citation information aggregated using \texttt{\href{https://flatironinstitute.github.io/software-citation-station/}{The Software Citation Station}} \citep{software-citation-station-paper,software-citation-station-zenodo}. Figure 1 was generated entirely, and Figure 5 in part, using Anthropic's Claude Opus 5.5 large language model.

The work by MR was supported by the U.S.\ Department of Energy through the Los Alamos National Laboratory (LANL). ROS acknowledges support from NSF awards AAG 1909534, 2206321 and PHY 2012057, 2309172. This research was made possible in part by the Information Science \& Technology Institute (ISTI) at LANL. This research used resources provided by LANL through the Institutional Computing (IC) program. Los Alamos National Laboratory is operated by Triad National Security, LLC, for the National Nuclear Security Administration of U.S.\ Department of Energy (Contract No.\ 89233218CNA000001). This document has been assigned LA-UR-26-30455.
  
\end{acknowledgements}

\begin{figure*}[ht!]
    \centering
    \includegraphics[width=\linewidth]{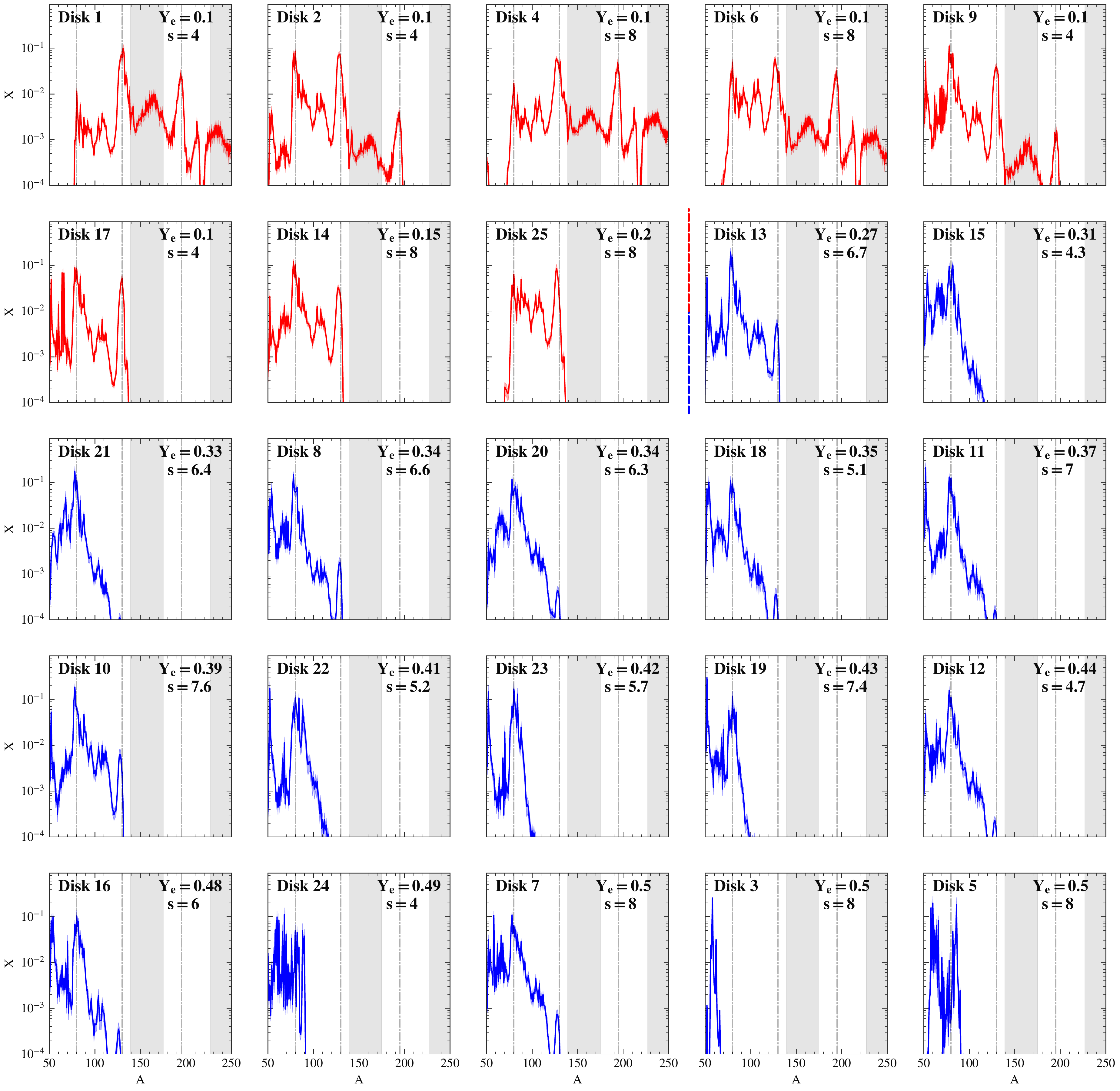}
    \caption{Mass fractions X as a function of mass number A for each disk, organized in order of increasing initial $Y_{\rm{e}}$ in the top right of each subplot (and then increasing $M_{\rm{bh}}$ for disks with equal $Y_{\rm{e}}$). The top-right label also includes the initial disk entropy $s$. The dashed vertical gray lines correspond to the first ($A \sim 80$), second ($A \sim 130$), and third ($A \sim 195$) $r$-process peaks, while the shaded gray regions indicate the lanthanides ($139 \lesssim A \lesssim 175$) and actinides ($227 \lesssim A \lesssim 262$). The nucleosynthesis yields presented here are mass-averaged across multiple $Y_{\rm{e}}$ bins as described in Section~\ref{sec:prism}. For this figure, tracers from each $Y_{\rm{e}}$ bin are randomly drawn, weighted by their masses, from a distribution assuming $v_r < 0.08c$ with $\theta$ left unconstrained.}
    \label{fig:vrlt}
\end{figure*}

\newpage

\begin{figure*}[ht!]
    \centering
    \includegraphics[width=\linewidth]{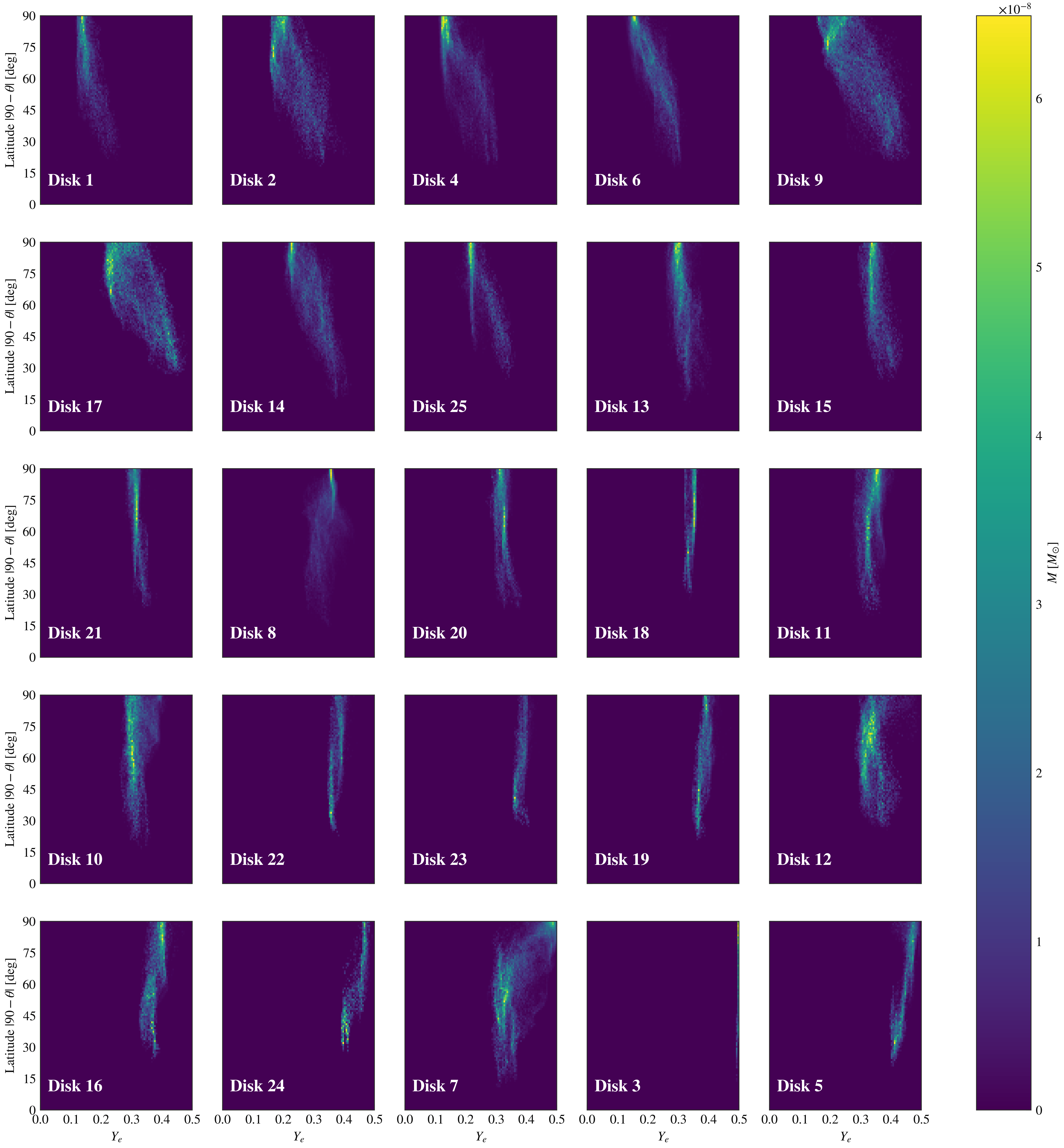}
    \caption{Mass histograms as a function of $Y_{\rm{e}}$ and latitude angle $90^\circ - \theta$. For this figure, tracers from each $Y_{\rm{e}}$ bin are randomly drawn, weighted by their masses, from a distribution assuming $v_r < 0.08c$ with $\theta$ left unconstrained.}
    \label{fig:vrlt_hist}
\end{figure*}

\newpage

\begin{figure*}[ht!]
    \centering
    \includegraphics[width=\linewidth]{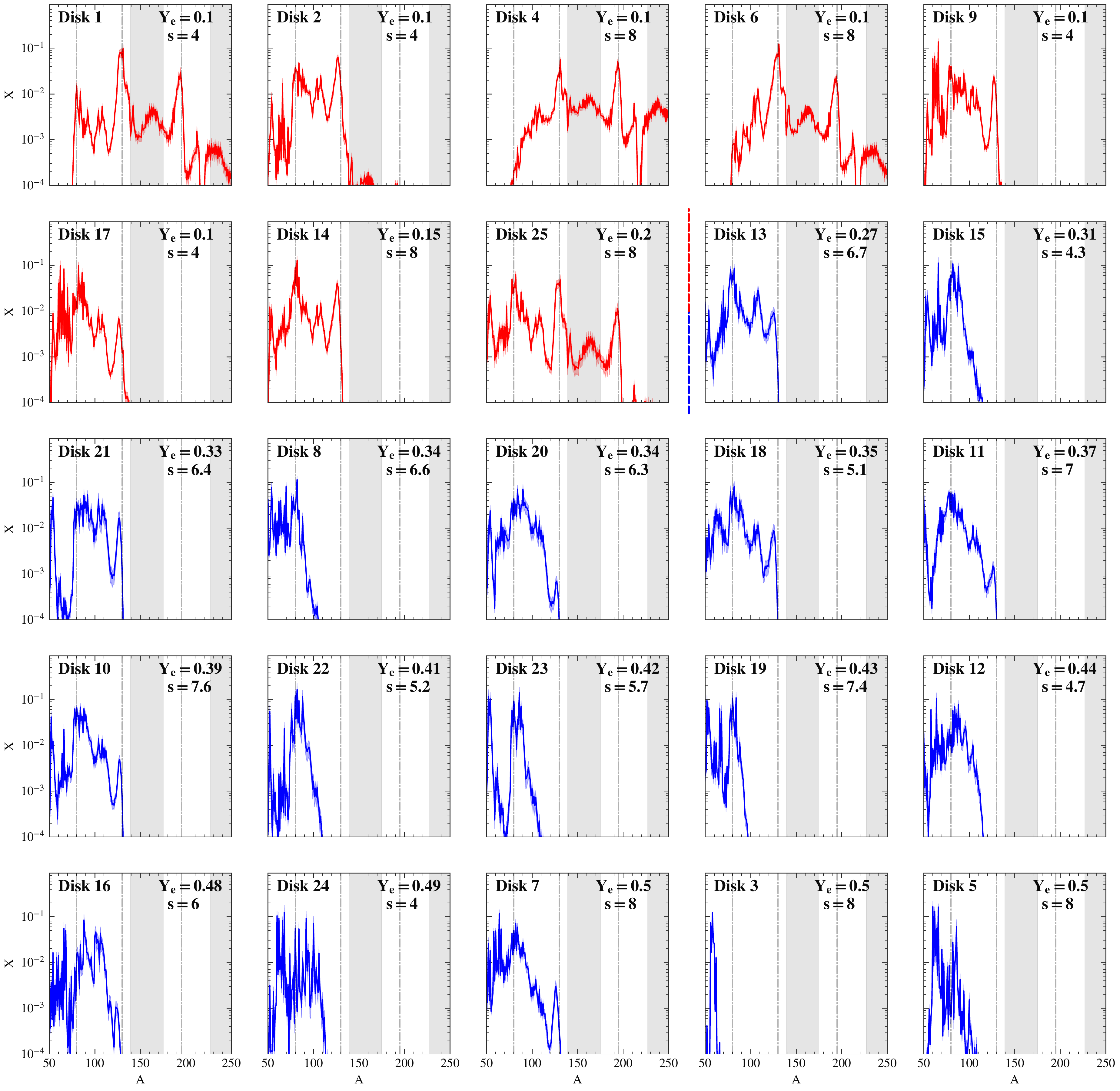}
    \caption{Same as Figure~\ref{fig:vrlt}, but with $v_r \geq 0.08c$ and $\theta$ unconstrained.}
    \label{fig:vrgt}
\end{figure*}

\newpage

\begin{figure*}[ht!]
    \centering
    \includegraphics[width=\linewidth]{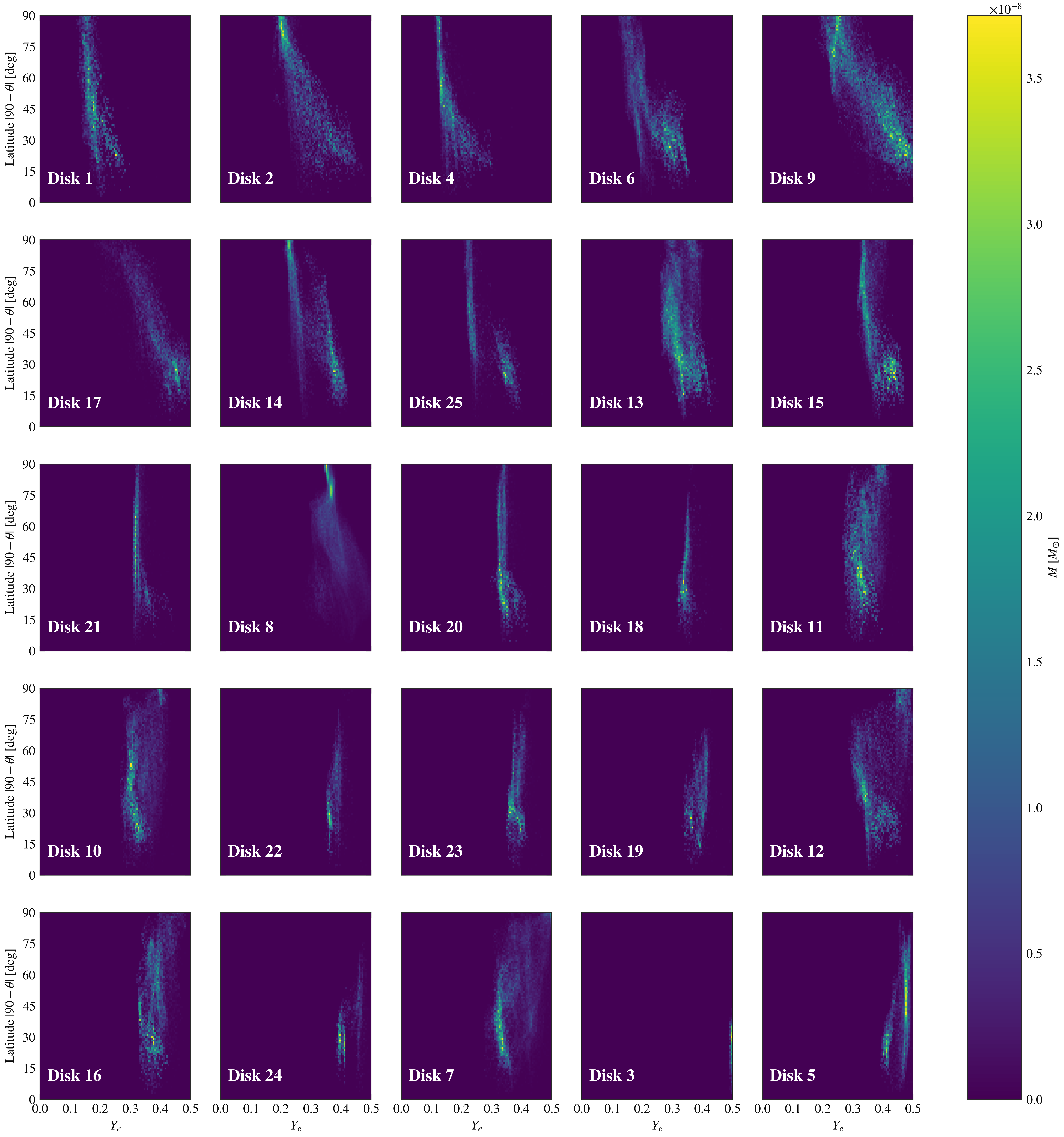}
    \caption{Same as Figure~\ref{fig:vrlt_hist}, but with $v_r \geq 0.08c$ and $\theta$ unconstrained.}
    \label{fig:vrgt_hist}
\end{figure*}

\newpage

\begin{figure*}[ht!]
    \centering
    \includegraphics[width=\linewidth]{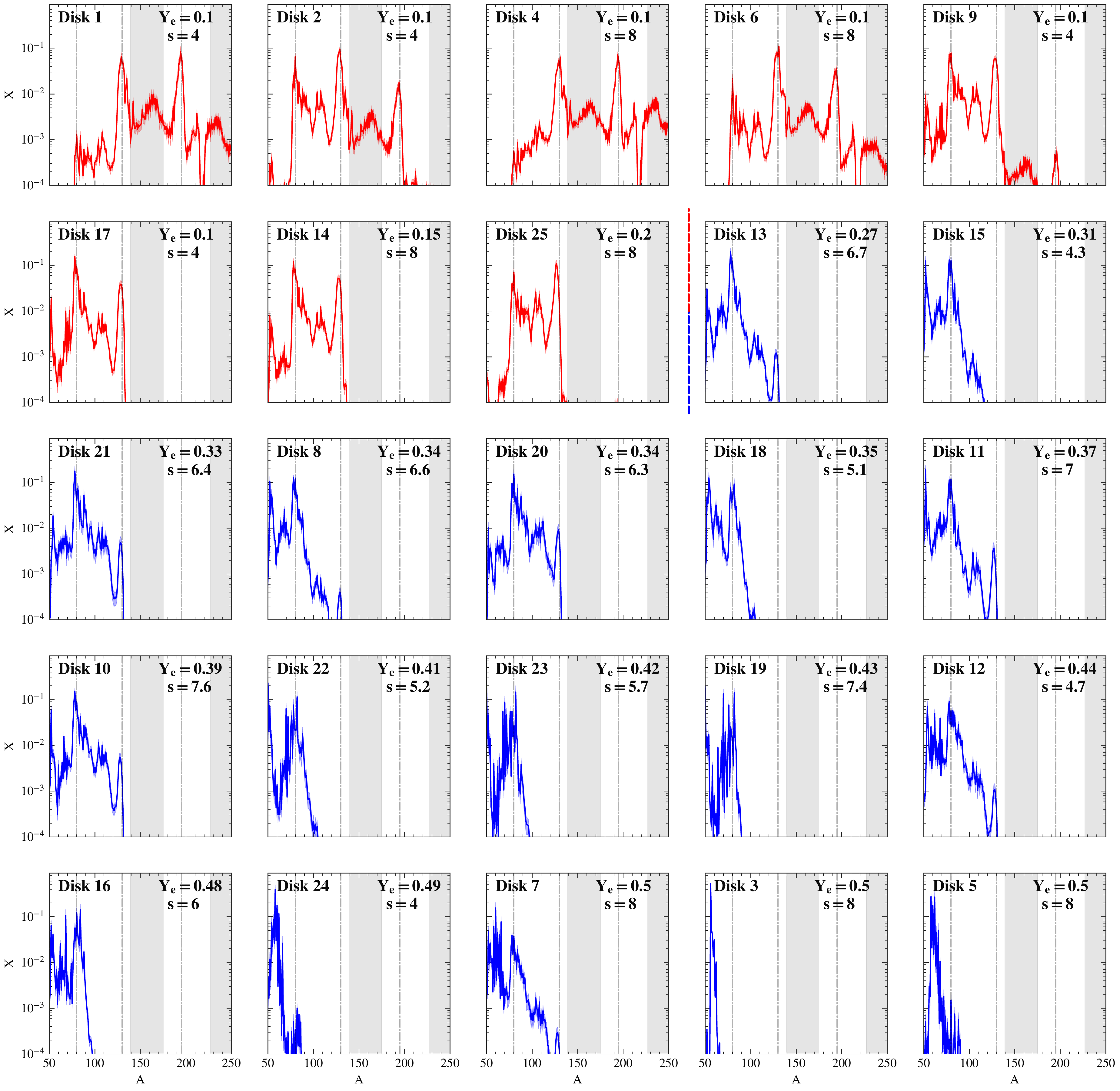}
    \caption{Same as Figure~\ref{fig:vrlt}, but with $\theta < 30^\circ$ and $v_r$ unconstrained.}
    \label{fig:thetalt}
\end{figure*}

\newpage

\begin{figure*}[ht!]
    \centering
    \includegraphics[width=\linewidth]{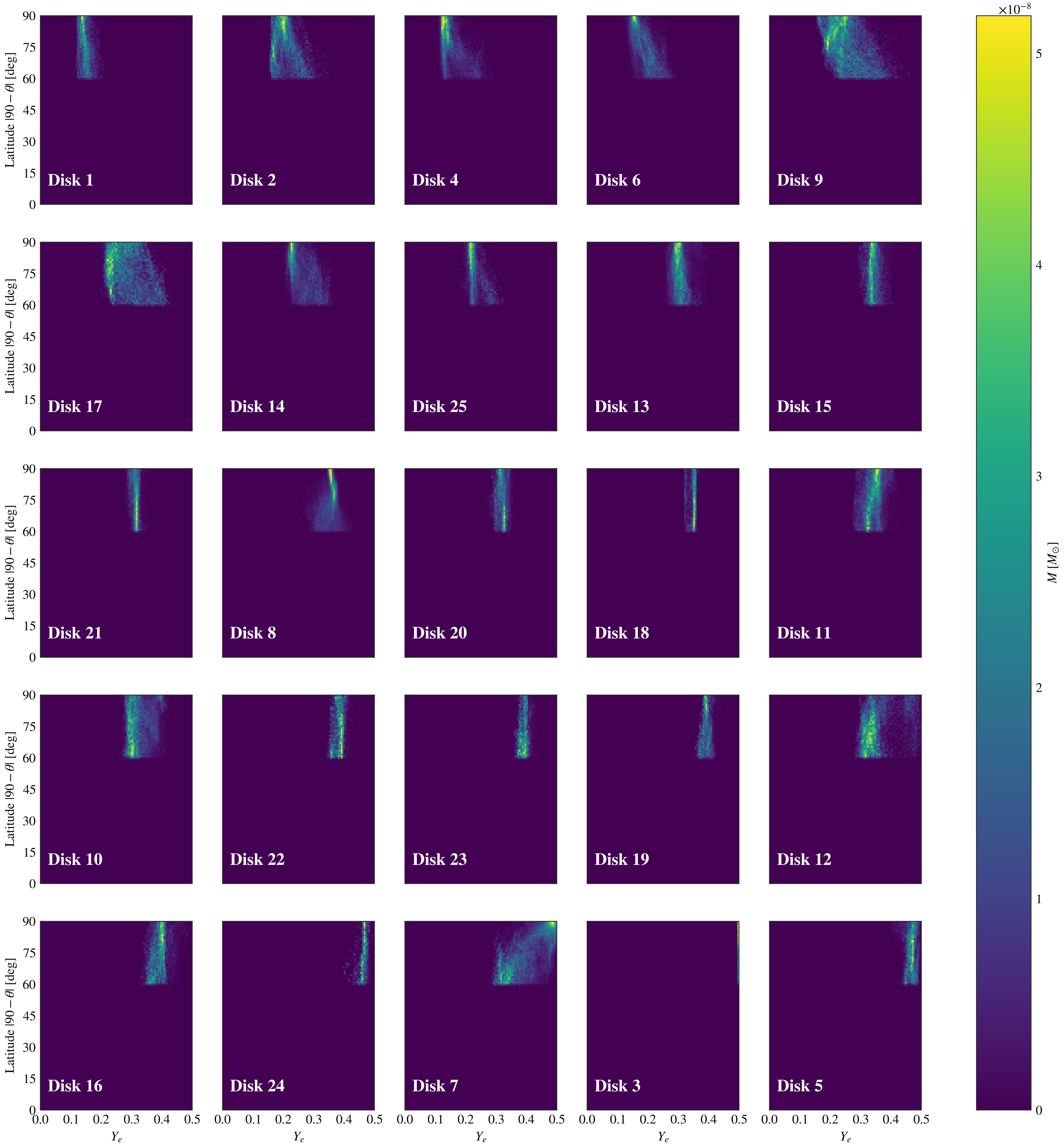}
    \caption{Same as Figure~\ref{fig:vrlt_hist}, but with $\theta < 30^\circ$ and $v_r$ unconstrained.}
    \label{fig:thetalt_hist}
\end{figure*}

\newpage

\begin{figure*}[ht!]
    \centering
    \includegraphics[width=\linewidth]{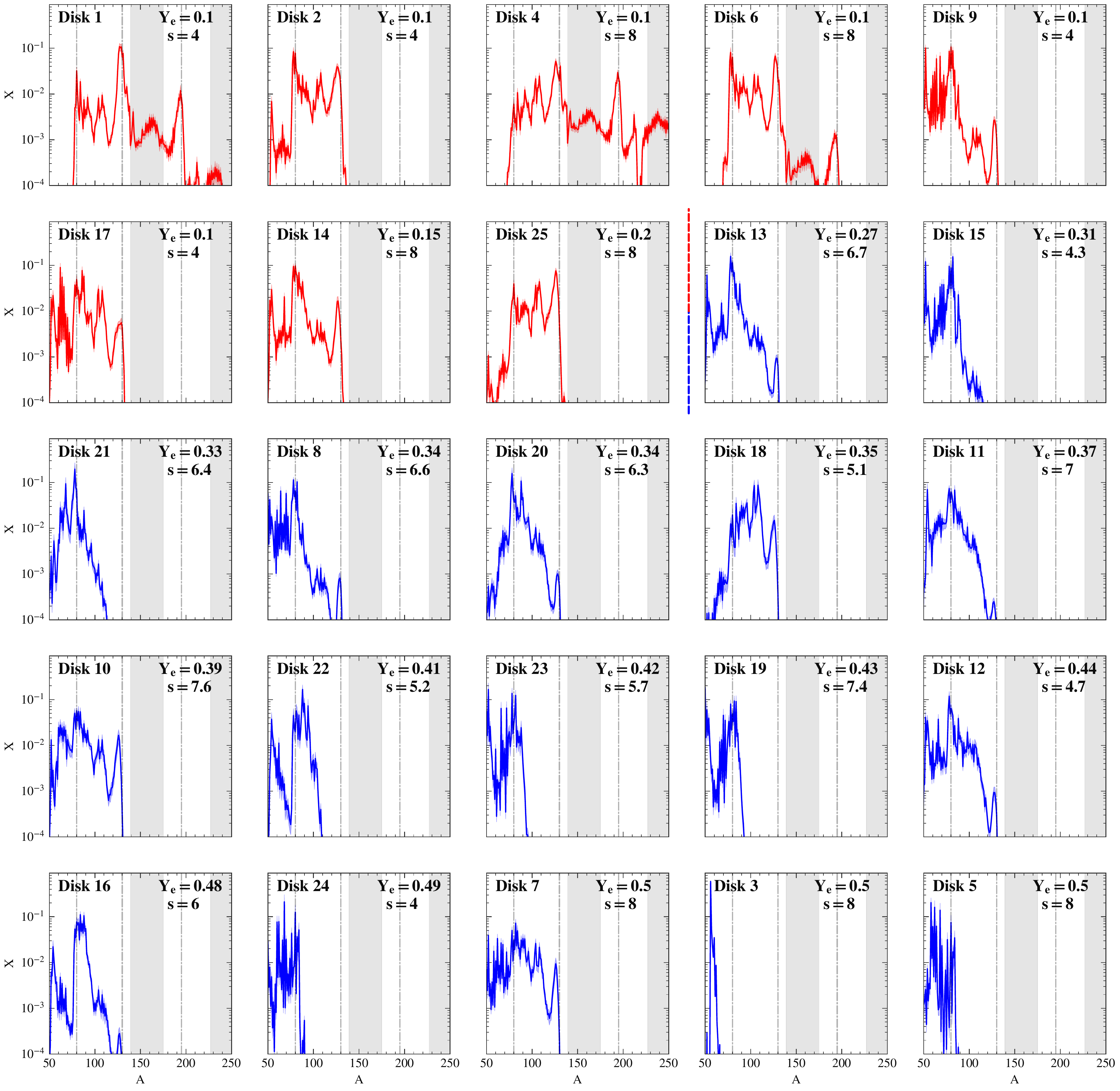}
    \caption{Same as Figure~\ref{fig:vrlt}, but with $\theta \geq 30^\circ$ and $v_r$ unconstrained.}
    \label{fig:thetagt}
\end{figure*}

\newpage

\begin{figure*}[ht!]
    \centering
    \includegraphics[width=\linewidth]{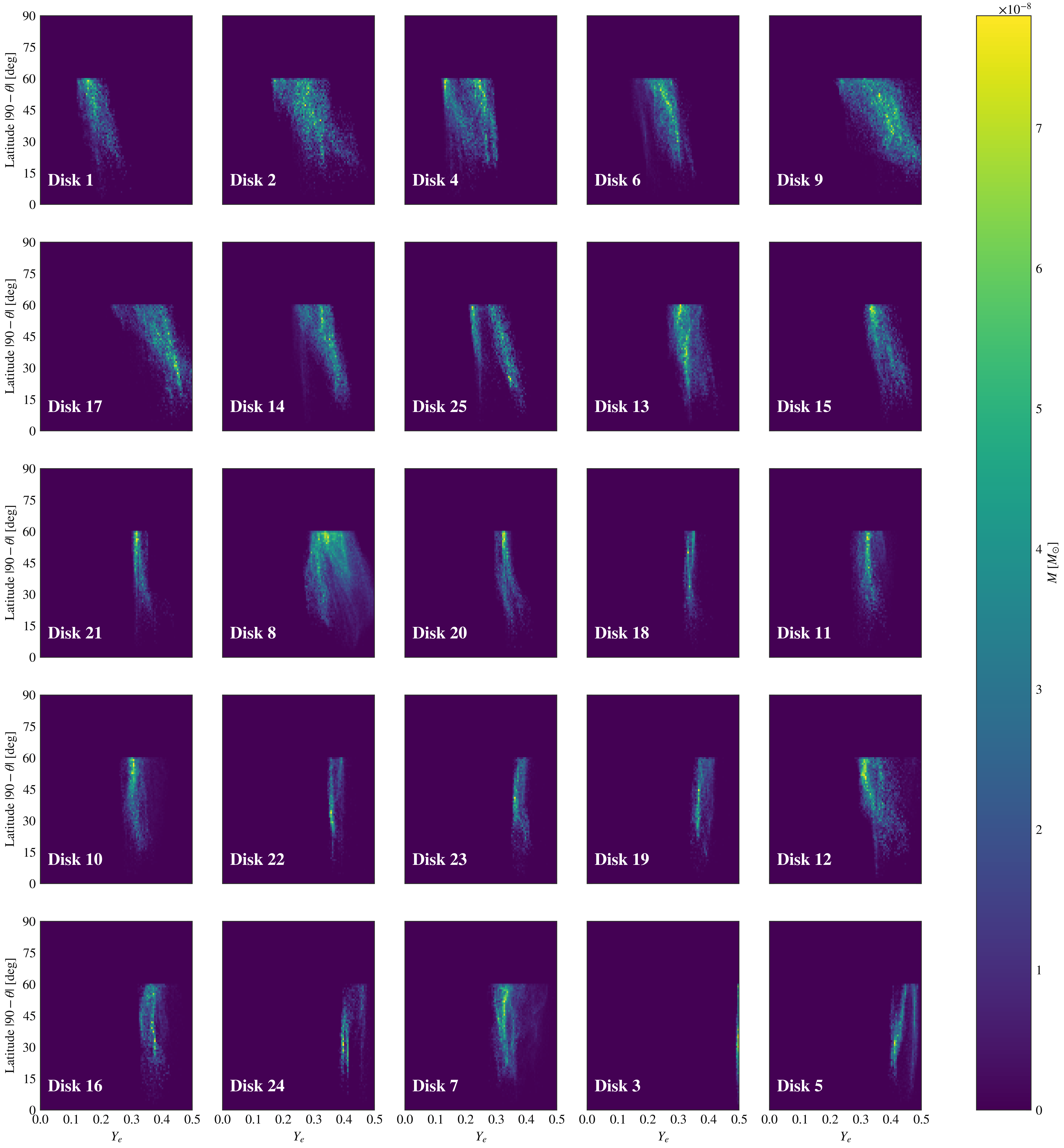}
    \caption{Same as Figure~\ref{fig:vrlt_hist}, but with $\theta \geq 30^\circ$ and $v_r$ unconstrained.}
    \label{fig:thetagt_hist}
\end{figure*}

\clearpage

\bibliographystyle{aasjournal}
\bibliography{refs}

\end{document}